\documentclass[aps,showpacs,twocolumn,floats,superscriptaddress,10pt]{revtex4-1}
\usepackage{dcolumn}
\usepackage[pdftex]{graphicx}

\usepackage[utf8]{inputenc}
\usepackage[UKenglish]{babel}
\usepackage{amsmath,amssymb,euscript,multirow} 
\usepackage{placeins}
 \usepackage{bbm}     
\newcommand\id{\ensuremath{\mathbbm{1}}}

\usepackage{ifthen}
\newcommand{\hideandshow}[1]{%
 \ifthenelse{\isundefined{\showme}}{}{#1}}
\newcommand{\showandhide}[1]{%
 \ifthenelse{\isdefined{\showme}}{}{#1}}
  
\RequirePackage[
   hyperindex,colorlinks,bookmarksnumbered,
   plainpages=true,pdfstartview=FitH]{hyperref}
\hypersetup{linkcolor=blue,urlcolor=blue,citecolor=blue}
\usepackage{hyperref}

\usepackage{color}
\usepackage{ulem}
\renewcommand{\emph}[1]{\textit{#1}}

\definecolor{darkblue}{rgb}{0,0,0.5}
\definecolor{darkgreen}{rgb}{0,0.5,0}
\definecolor{darkred}{rgb}{.7,0,0}
\definecolor{purple}{rgb}{0.5,0,0.6}
\definecolor{orange}{rgb}{1,0.5,0}
\definecolor{grey}{rgb}{.6,.6,.6}
\definecolor{lightpink}{rgb}{1,0.7,0.75}
\definecolor{pink}{rgb}{1,0.4,0.58}
\definecolor{deeppink}{rgb}{1,0.08,0.58}

\newcommand{\comment}[1]{{\color{blue}{[\textbf{Comment:} \textit{
        #1}]}}}

\newcommand{\Eq}[1]{Eq.~(\ref{#1})}

\newcommand{\Sec}[1]{Sec.~\ref{#1}}

\newcommand{\Fig}[1]{Fig.~\ref{#1}}

\newcommand{\Ref}[1]{Ref.~[\onlinecite{#1}]}
\newcommand{\Refs}[1]{Refs.~[\onlinecite{#1}]}

\newcommand{\pdag}{{\phantom{\dagger}}}

\newcommand{\qqph}{{\phantom{\qquad}}}
\newcommand{\bathspectrum}{{\rm bath}}
\newcommand{\mflavor}{m_{\rm f}}
\newcommand{\slow}{{S}}
\newcommand{\fast}{{F}}
\newcommand{\BATH}{{B}}
\newcommand{\bath}{{\rm bath}}

\newcommand{\imp}{{\rm imp}}
\newcommand{\cut}{{\rm c}}
\newcommand{\dyn}{{\rm dyn}^\pdag}
\newcommand{\stat}{{}}
\newcommand{\exact}{{\rm exact}^\pdag}

\newcommand{\alphac}{{\alpha_\cut}}

\newcommand{\DHO}{{\rm DHO}}
\newcommand{\SBM}{{\rm SBM}}
\newcommand{\CFE}{{\rm CFE}}

\newcommand{\OWC}{{\rm OWC}}

\newcommand{\SWC}{{\rm SWC}}
\newcommand{\RWC}{{\rm RWC}}
\newcommand{\UV}{{\rm uv}}
\newcommand{\Czero}{{\rm C0}^\pdag}
\newcommand{\Cone}{{\rm C1}^\pdag}
\newcommand{\Ctwo}{{\rm C2}^\pdag}
\newcommand{\CZero}{{\rm C0}}
\newcommand{\COne}{{\rm C1}}

\newcommand{\TT}{{\rm T}}
\newcommand{\MM}{{\rm M}}
\newcommand{\GG}{{\rm G}}
\newcommand{\WW}{{\rm W}}
\newcommand{\VV}{{\rm V}}
\newcommand{\PP}{{\rm P}}

\newcommand{\BTV}{{\rm BTV}}

\newcommand{\TBM}{{\rm TBM}}

\newcommand{\Ham}{\mathcal{H}}
\newcommand{\G}{\mathcal{G}}
\newcommand{\selfE}{\Sigma}
\newcommand{\uselfE}{\underline{\Sigma}}
\newcommand{\A}{\mathcal{A}}
\newcommand{\Omegar}{\Omega_{\rm r}}
\newcommand{\w}{\omega}

\newcommand{\uGamma}{\underline{\Gamma}}

\newcommand{\ulambda}{\underline{\lambda}}

\newcommand{\uone}{\underline{1}}
\newcommand{\bnu}{\bar \nu}
\newcommand{\ub}{\underline{b}}
\newcommand{\uf}{\underline{f}}
\newcommand{\ut}{\underline{t}}
\newcommand{\uvarepsilon}{\underline{\varepsilon}}
\newcommand{\uG}{\underline{\mathcal{G}}}
\newcommand{\uA}{\underline{\mathcal{A}}}
\newcommand{\uSigma}{\underline{\Sigma}}

\newcommand{\uu}{\underline{u}}
\newcommand{\ud}{\underline{d}}

\newcommand{\uw}{\underline{w}}

\begin{document}

\title{Open Wilson chains for quantum impurity models: Keeping track of 
all bath modes}

\author{B. Bruognolo}
\affiliation{Physics
  Department, Arnold Sommerfeld Center for Theoretical Physics and
  Center for NanoScience, Ludwig-Maximilians-Universit\"at M\"unchen,
  D-80333 M\"unchen, Germany}
\affiliation{Max-Planck-Institut f\"ur Quantenoptik, 
Hans-Kopfermann-Str. 1, D-85748 Garching, Germany}
\author{N.-O. Linden}
\affiliation{Physics
  Department, Arnold Sommerfeld Center for Theoretical Physics and
  Center for NanoScience, Ludwig-Maximilians-Universit\"at M\"unchen,
  D-80333 M\"unchen, Germany}
\author{F. Schwarz}
\affiliation{Physics
  Department, Arnold Sommerfeld Center for Theoretical Physics and
  Center for NanoScience, Ludwig-Maximilians-Universit\"at M\"unchen,
  D-80333 M\"unchen, Germany}
\author{S.-S. B. Lee}
\affiliation{Physics
  Department, Arnold Sommerfeld Center for Theoretical Physics and
  Center for NanoScience, Ludwig-Maximilians-Universit\"at M\"unchen,
  D-80333 M\"unchen, Germany}
\author{K. Stadler}
\affiliation{Physics
  Department, Arnold Sommerfeld Center for Theoretical Physics and
  Center for NanoScience, Ludwig-Maximilians-Universit\"at M\"unchen,
  D-80333 M\"unchen, Germany}
\author{\mbox{A. Weichselbaum}} 
\affiliation{Physics
  Department, Arnold Sommerfeld Center for Theoretical Physics and
  Center for NanoScience, Ludwig-Maximilians-Universit\"at M\"unchen,
  D-80333 M\"unchen, Germany}
\author{M. Vojta} \affiliation{Institut
  f\"ur Theoretische Physik, Technische Universit\"at Dresden, D-01062
  Dresden, Germany}
\author{F. B. Anders} \affiliation{Lehrstuhl f\"ur
  Theoretische Physik II, Technische Universit\"at Dortmund, D-44221
  Dortmund, Germany}
\author{J. von Delft}  \affiliation{Physics
  Department, Arnold Sommerfeld Center for Theoretical Physics and
  Center for NanoScience, Ludwig-Maximilians-Universit\"at M\"unchen,
  D-80333 M\"unchen, Germany}


\begin{abstract}
  When constructing a Wilson chain to represent a quantum impurity
  model, the effects of truncated bath modes are neglected.  We show
  that their influence can be kept track of systematically by
  constructing an ``open Wilson chain'' in which each site is coupled
  to a separate effective bath of its own. As a first application, we use the method to
  cure the so-called mass-flow problem that can arise when using
  standard Wilson chains  to treat impurity models
  with asymmetric bath spectral functions at finite temperature.  We
  demonstrate this for the strongly sub-Ohmic spin-boson model at
  quantum criticality where we directly observe the flow towards a
  Gaussian critical fixed point.
\end{abstract}

\maketitle

A quantum impurity model describes a discrete set of degrees of
freedom, the ``impurity'', coupled to a bath of excitations. For an
infinite bath this is effectively an \textit{open} system. However,
the most powerful numerical methods for solving such models, Wilson's
numerical renormalization group (NRG) \cite{Wilson1975,Bulla2008} and
variational matrix-product-state (VMPS) generalizations thereof
\cite{Saberi2008,Weichselbaum2009,Pizorn2012,Guo2012,*Bruognolo2014},
actually treat it as \textit{closed}: The continuous bath is replaced
by a so-called Wilson chain, a finite-length tight-binding chain whose
hopping matrix elements $t_n$ decrease exponentially with site number
$n$, ensuring energy-scale separation along the chain.
This works well for numerous applications, ranging from transport
through nanostructures \cite{Borda2003,Kretinin2011} to  
impurity solvers for dynamical mean-field theory  
\cite{Bulla1999,Pruschke2005,Stadler2015b}. However, replacing an open
by a closed system brings about finite-size effects.  Wilson
  himself had anticipated that the effect of bath modes neglected
  during discretization might need to be included perturbatively ``to
  achieve reasonable accuracy'', but concluded that ``this has proven
  to be unnecessary'' for his purposes (see p.~813 of
  Ref.~\cite{Wilson1975}). By now, it is understood that finite-size
  effects often do matter.   They hamper the 
treatment of dissipative effects \cite{Rosch2012}, e.g., in the
context of nonequilibrium transport \cite{Anders2008a} and
equilibration after a local quench  
\cite{Anders2005,*Tureci2011,*Latta2011}.  Moreover, even in
equilibrium, they may cause errors when computing the bath-induced
renormalization of impurity properties
\cite{Vojta2005,Vojta2009,Vojta2010}.  Indeed, finite-size issues
  constitute arguably the most serious conceptual limitation of
  approaches based on Wilson chains.   
 
Here we set the stage for controlling finite-size effects by
constructing ``open Wilson chains'' (OWCs) in which each
site is coupled to a bath of its own. The resulting open system implements
energy-scale separation in a way that, in contrast to standard Wilson chains (SWC), fully keeps track of all
bath-induced dissipative and renormalization effects. 
The key step involved in any renormalization group (RG) approach, namely integrating out degrees of freedom at one energy scale to obtain a renormalized description at a lower scale, can then be performed more carefully than for SWCs. 
We illustrate this  by focusing on renormalization
effects, leaving a systematic treatment of dissipative effects on OWCs for the future.

A SWC is constructed by logarithmically discretizing the bath and
tridiagonalizing the resulting discrete bath Hamiltonian to obtain a
tight-binding chain, with the impurity coupled to site $n=0$
\cite{Wilson1975,Bulla2008}.  Properties at temperature $T$ are
calculated using a chain of finite length
$N_T$, 
chosen such that its smallest energy scale matches the temperature
$t_{N_T} \simeq T$ ($k_B=1$). However, since sites $n> N_T$ are
neglected, the contribution of the corresponding truncated bath modes
(TBMs) to the renormalization of impurity properties is missing
\cite{Vojta2010}.  For example, for a local level linearly coupled to
a bath with an asymmetric bath spectrum, this coupling generates a
physical shift in the level energy. 
When this shift is computed using a SWC of length $N_T$, the
  re\-sult contains a temperature-dependent error. Hence, the use of
  SWCs generically leads to qualitative errors in the temperature
  dependence of renormalized model parameters, called the ``mass-flow
  problem'' \cite{Vojta2009,Vojta2010}.  Quantitative errors 
  persist even for $T \! \to \! 0$, when $N_T \! \to \! \infty$,
  because constructing a SWC actually involves neglecting TBMs at \textit{every} site.

The mass-flow problem is particularly serious 
when targeting a quantum critical point, 
where it causes errors for critical exponents describing
finite-temperature properties at the critical point.
This has been studied in some detail for the dissipative harmonic
oscillator (DHM) and the sub-Ohmic spin-boson model (SBM). For both,
SWCs are unable to even qualitatively describe the temperature
dependence of the local susceptibility $\chi(T)$ at criticality
\cite{Vojta2005,Vojta2009,Vojta2010}.  Both involve Gaussian
criticality of $\phi^4$ type and hence a bosonic mode whose excitation
energy vanishes at the critical fixed point. The finite-temperature RG flow in
its vicinity cannot be correctly described using finite-length SWCs
because the erroneous mass dominates over physical interaction
effects.
Summarizing, methods based on SWCs produce systematic quantitative
errors for all impurity problems with asymmetric baths, and they fail
even qualitatively in addressing Gaussian criticality and other
phenomena with zero modes.

Here we show that these issues \textit{can} be addressed using OWCs:
The bath coupled to each site of the OWC induces an energy shift for
that site that can be computed \textit{exactly} and used to define a
``renormalized Wilson chain'' (RWC).  The ground-state properties of a
RWC of length $N_T$ mimic the finite-$T$ properties of the original
model in a way that is free from mass-flow problems.  We demonstrate
this explicitly by using VMPS techniques \cite{Bruognolo2014} on RWCs
to compute $\chi(T)$ for the DHO and SBM. We also compute the
energy-level flow of the SBM; it unambiguously reveals
flow towards a Gaussian fixed point with a dangerously
irrelevant interaction term.

\begin{figure}[t]
\includegraphics[width=0.99\linewidth]{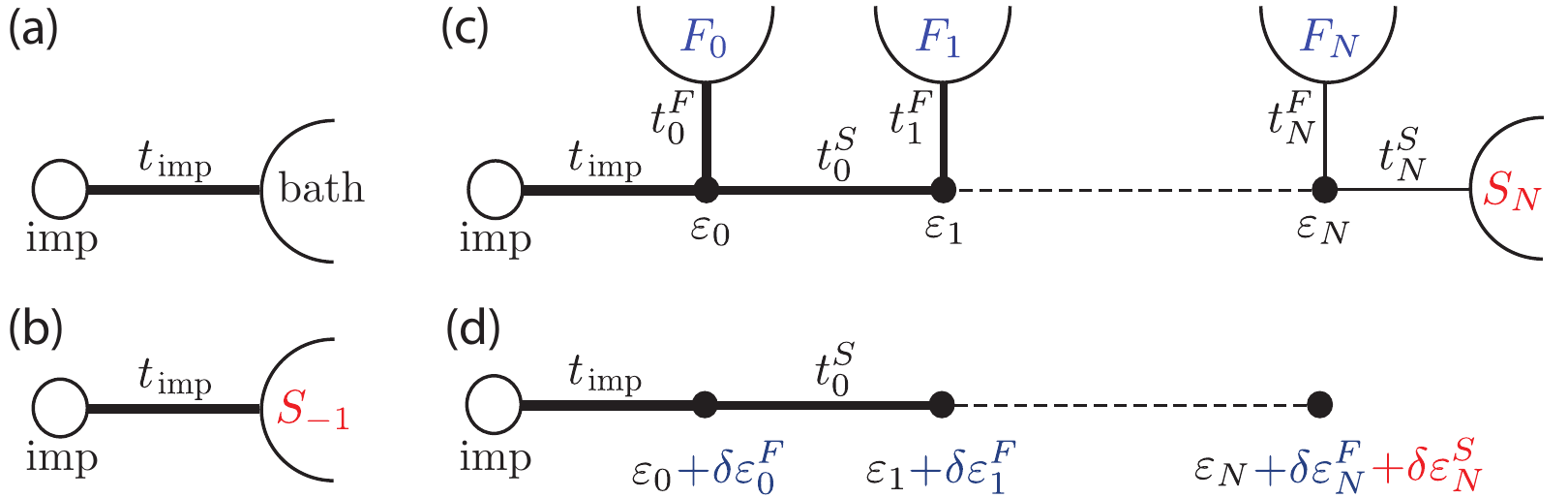}
\vspace{-6mm}
\caption{(a) Impurity model. (b) Initialization. (c) 
Open Wilson chain (OWC).   (d) Renormalized Wilson chain (RWC).
\label{fig:OWC}} \vspace{-8mm}
\end{figure}   
  

\textit{Model.} We consider a generic single-band impurity model with Hamiltonian
$\mathcal{H} = \mathcal{H}^\imp[ b^\dag t_\imp ] + \mathcal{H}^\bath$,
where $\mathcal{H}^\bath$ describes the bath, and $\mathcal{H}^\imp$
the impurity and its coupling to the bath via normalized bath
operators $b^\dag$ and $b$, with coupling constant $t_\imp$
[Fig.~\ref{fig:OWC}(a)].  The free ($t_\imp\!=\!0$) dynamics of
$b^\dag$, generated by $\Ham^\bath$, is encoded in the free retarded
correlator
$\G^\bath(\omega) = \langle \! \langle b | b^\dagger \rangle \! \rangle_\omega$, 
which is uniquely characterized by its spectral function
$\A^\bath(\omega) = - \frac{1}{\pi} {\rm Im} \, \G^\bath(\omega)$.
The impurity dynamics is therefore fully determined once 
$\Ham^\imp$
and the ``bath spectrum'',
$\Gamma^\bath (\omega) = |t_\imp|^2 \A^\bath (\omega)$,
have been specified.

\textit{Continued-fraction expansion.} 
 One well-known way of mapping an impurity model to a chain is to
  iteratively construct a conti\-nued-fraction expansion (CFE) for
  $\G^\bath$
  \cite{Grosso1985,*Foulkes1986,*Gagliano1987,*Si1994,*Hallberg1995}.
  Our main idea is to do this in a way that \textit{zooms in on
    low energies}  without discarding high-energy information.  Our
  construction involves a sequence of retarded correlators
  $\G_n^X(\omega)$, with $X \!=\! S$ or $F$, describing the effective
  ``slow'' (low-energy) or ``fast'' (high-energy) bath modes of
  iteration step $n$, with spectral functions
  $\A_n^X(\omega) = -\frac{1}{\pi} \text{Im} \, \G_n^X(\omega)$ having
  unit weight, $\int d \omega \A_n^X (\omega) = 1$.
  We initialize our CFE construction with 
  $\G_{-1}^\slow \! = \! \G^\bath$ [Fig.~\ref{fig:OWC}(b)]. Starting
  with $n\! =\! 0$, we iteratively use $\G^\slow_{n-1}$, describing
  the low-energy modes of the previous iteration, as input to define a
  new retarded correlator $\G_n$ and its retarded self-energy
  $\selfE_n$, 
\begin{align}
\label{eq:defineretarded}
\G_n(\omega) = \G^\slow_{n-1}(\omega) = 
1/\left[\omega - \varepsilon_n
- \selfE_n(\omega)\right] , 
\end{align}
with $\varepsilon_n = \int d \omega \, \omega \A_n(\omega)$
\cite{supplement}. Then we  split this self-energy into
low- and high-energy parts by writing it as 
\begin{eqnarray}
  \label{eq:shortendedsplit-bath}
\selfE_n(\omega) = 
\selfE_n^\slow (\omega) + \selfE_n^\fast (\omega) , 
\quad \selfE_n^X (\omega) = |t_n^X|^2 \G_n^X(\omega) .
\phantom{.} \quad 
\end{eqnarray} 
Here the corresponding retarded correlators $\G_n^{\slow/\fast}$ are defined
 by choosing their rescaled spectral functions,
 $|t_n^{\slow/\fast}|^2 \A_n^{\slow/\fast}$,
to represent the low- and high-energy 
parts of $\Gamma_n(\omega) = -\frac{1}{\pi} \text{Im} \Sigma_n$, 
with $t_n^X$ chosen such that $\A^X_n$ has unit weight
(see Sec.~S-1~A of Ref.~\cite{supplement} for details). 
To be explicit, we write $\Gamma_n = \Gamma_n^{\slow} + \Gamma_n^{\fast}$, with $\Gamma^X_n (\omega) = w^X_n(\omega) \Gamma_n (\omega)$. The splitting functions $w_n^{\slow/\fast} (\omega)$ are defined
on the support of $\Gamma_n$, take values in the interval $[0,1]$, satisfy
$w_n^\slow (\omega) + w_n^\fast (\omega) = 1$, and have weight
predominantly at low/high energies. Then we write the split bath spectra as
$\Gamma^{X}_n (\omega) = |t_n^X|^2 \A^X_n (\omega)$, 
with ``couplings'' $t_n^X$ chosen as 
$ |t_n^X|^2 = \int d \omega \, \Gamma^X_n(\omega)$,
and define
new retarded correlators via
$\G_n^X(\omega) = \! \int d \bar \omega\, \frac{\A_n^X(\bar
  \omega)}{\omega - \bar \omega + i 0^+}$,  also fixing $\Sigma^X_n(\omega)$ via \Eq{eq:shortendedsplit-bath}.

Iterating, using $\G_n^\slow$ as input to compute new correlators
$\G_{n+1}^X$ while retaining the self-energy
$\Sigma_n^\fast$, we obtain a sequence of exact CFE
re\-pre\-sen\-ta\-tions for $\G^\bath$.  That of depth 2, e.g., reads
\vspace{-1mm} 
\begin{align*} 
\G^\bath (\omega) 
& = 
  \frac{1}{\omega- \varepsilon_0  - \! \selfE_0^F  (\omega) - 
      \frac{\left|t^S_0\right|^{2^{\color{white}
            2}} 
      }{\omega-\varepsilon_1 - 
\selfE_1^F(\omega) - \frac{|t_1^\slow|^2}{\omega - \varepsilon_2 - 
\selfE_2 (\omega)}}} .  
\end{align*} 
To ensure energy-scale separation, we choose  
$\A^X_n(\omega)$ such that the CFE parameters decrease
monotonically,   
$\text{max}\{|\varepsilon_n|,|t^\slow_n|\} \leqslant
\text{max}\{|\varepsilon_{n-1}|,|t^\slow_{n-1}|\}/\Lambda$,
with $\Lambda \!>\! 1$ \footnote{If $\Gamma^\bath (\omega)$ has
  power-law form, $\propto \! \omega^s$, we can achieve this by taking the
  support of $\A^{\slow}_n $ and $\A^{\fast}_n$ to \textit{partition}
  that of $-\frac{1}{\pi}\text{Im}\, \Sigma_n$ 
into low- and high-energy regimes  \cite{supplement}.}.

\textit{Open Wilson chain.} We now use the CFE data
  ($\varepsilon_n, t_n^X, \G_n^X$) to represent the 
 original bath in terms
  of a chain with $N+1$ sites, each coupled to a bath of its own,
and site 0 coupled to the impurity (site $-1$)
[Fig.~\ref{fig:OWC}(c)]. 
  This OWC is constructed such that the free ($t_\imp=0$)
  correlator of site 0 is exactly equal to the depth-$N$ CFE
found above, i.e.\ $\G_0 = \G^\bath$, implying
that the chain and original bath have the same
effect on the impurity.

The key point is that each CFE step of writing
$\G^\slow_{n-1}(\omega)$ in the form
$\G_n(\omega) = 1/[\omega - \varepsilon_n - \Sigma_n(\omega)]$ can be
implemented on the level of the Hamiltonian: It corresponds to
replacing the bath represented by $\G^\slow_{n-1}$, say $S_{n-1}$, by
a new site $n$, with energy $\varepsilon_n$ and normalized site
operators $f^\dag_n$ and $f_n^\pdag$, 
which is linearly coupled to a new bath that
generates the self-energy $\Sigma_n$. In the present case, the latter
is split into low- and high-energy contributions,
$\Sigma_n^\slow + \Sigma_n^\fast$.  We can generate these by linearly
coupling the new site with couplings $t_n^\slow$ and $t_n^\fast$ to
\textit{two} new baths, say $S_n$ and $F_n$, via normalized bath
operators $b_{\slow n}^\dagger$, $b_{\slow n}^\pdag$ 
and $b_{\fast n}^\dagger$, $b_{\fast n}^\pdag$, that are
governed by bath Hamiltonians $\Ham_n^X$ chosen such that
$\langle \! \langle b_{Xn}^\pdag|b^\dagger_{Xn}
  \rangle \!  \rangle_\omega$ equals the $\G_n^X(\omega)$ found above  
  (see Sec.~S-1~A of Ref.~\cite{supplement} for details).  For the next
  iteration, we retain the fast bath $F_n$, but replace the slow bath
  $S_n$ by a new site $n+1$ coupled to new baths $S_{n+1}$ and
  $F_{n+1}$, etc.  This leads to replacing $\Ham$ by 
$\Ham^\OWC_N = \Ham^\SWC_N + \Ham^\TBM_N $, with
\vspace{-1mm} 
\begin{eqnarray}
  \label{eq:shortendedH-OWC}
  \Ham^\SWC_N
& = &   \Ham^\imp_f +
 \! \sum_{n=0}^N \varepsilon_n f^\dagger_{n} f^\pdag_{n} +
 \! \sum_{n = 0}^{N-1} (  f^\dagger_{n+1} t^S_n   f^\pdag_{n}
+ {\rm H.c.}) , \nonumber
\\
  \Ham^\TBM_N
 & = &
 \sum_{n = 0}^{N} ( b^\dagger_{\fast n} t^\fast_n f^\pdag_{n} + {\rm H.c.})
+ \sum_{n=0}^N \mathcal{H}^\fast_n
\\ \nonumber
& & +   ( b^\dagger_{\slow N} t^\slow_N f^\pdag_{N} + {\rm H.c.})
+ \mathcal{H}^\slow_N \; ,
\end{eqnarray}
and $\Ham^\imp_f = \Ham^\imp[ f^\dagger_0 \, t_\imp] $. 
This chain Hamiltonian is depicted schematically in Fig.~\ref{fig:OWC}(c). 
$\Ham^\SWC_N$ has the structure of a SWC, while $\Ham^\TBM_N $
describes the couplings to all fast baths $F_{n \leqslant N}$, and of
the last site $N$ to its slow bath
$S_N$.  
These ``fast and last slow'' baths $F_n$ and
$S_{N}$ constitute TBMs, since a SWC neglects them.  By instead using
an OWC, we can keep track of their influence, namely to shift, mix and
broaden the eigenstates of those subchains to which they couple. Equation (\ref{eq:shortendedH-OWC}), which represents an impurity model in terms of a Wilson chain that still is a fully open system, is the first main result of this Rapid Communication.


\begin{figure}[t]
  \centering
\includegraphics[width=.99\linewidth]{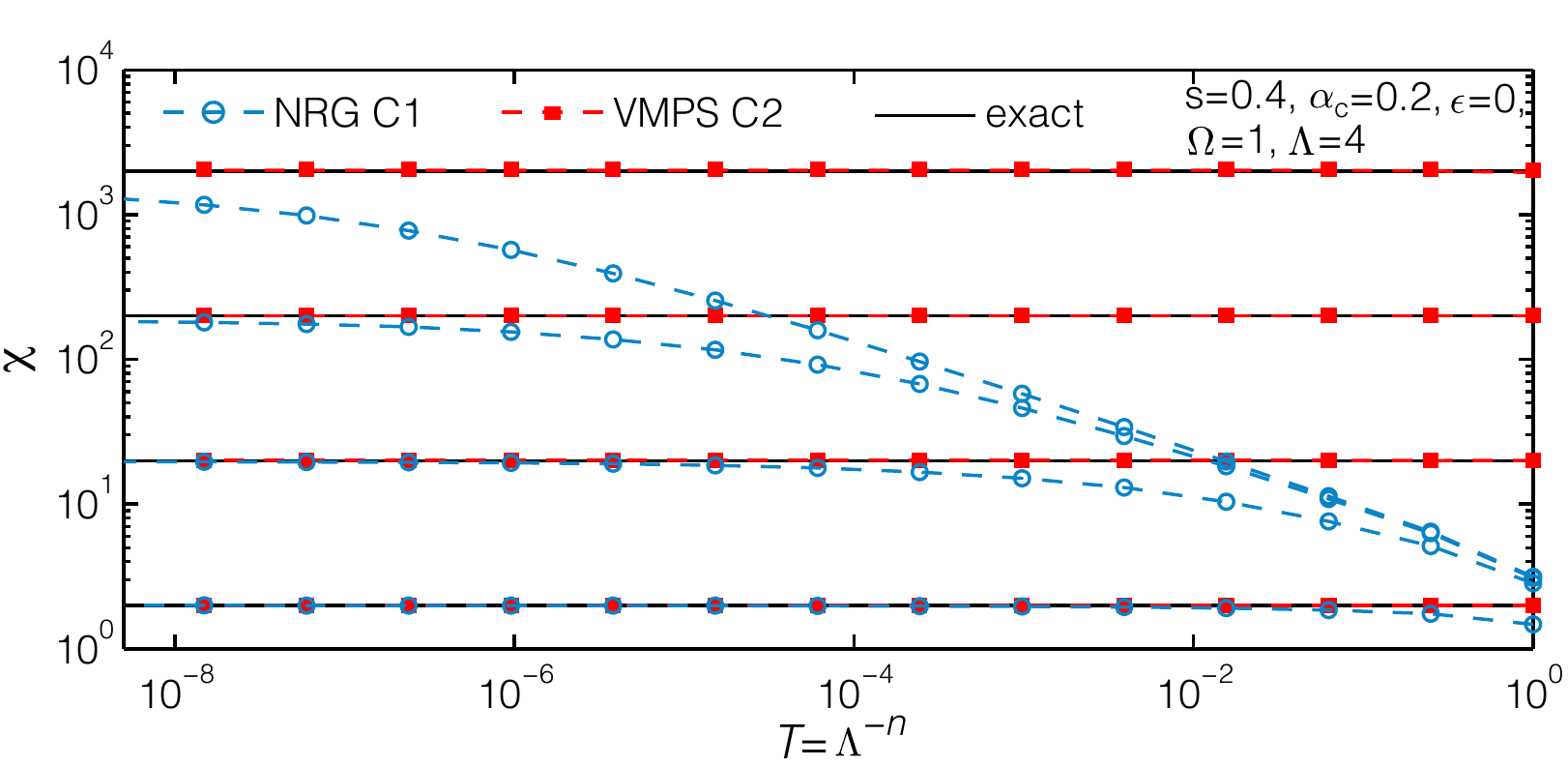} \vspace{-6mm}
\caption{DHO susceptibility $\chi(T)$ as function of
temperature, computed by NRG on C1-RWCs (circles)
and by VMPS on C2-RWCs (squares),
for  $\alpha=0.1$, 0.19, 0.199, and  0.1999 (from bottom to top).
Solid lines show exact results. }
\vspace{-6mm}
  \label{Fig.SuscCompare}
\end{figure}

\textit{Renormalized Wilson chain.} For concrete numerical
calculations, we need to approximate an OWC by a RWC
that can be treated using standard NRG or VMPS methods, while still
including information about the TBMs.
To this end, we replace $\Ham^\OWC$ by $\Ham^\RWC$
[Fig.~\ref{fig:OWC}(d)], a Hamiltonian of the same form as
  $\Ham^\SWC$ (without fast or last baths), but with each on-site
  energy $\varepsilon_n$ shifted to
\begin{eqnarray}
  \tilde \varepsilon_n = \varepsilon_n +
  \delta \varepsilon^\fast_n + \delta_{nN} \, \delta \varepsilon^\slow_N \; , \quad
  \delta\varepsilon^X_n={\rm Re}\bigl[\selfE_n^X(0) \bigr]
  \!  . \qqph
\label{eq.EnergyShift}  
\end{eqnarray} 
For the CFE of $\G^\bath = \G^\slow_{-1} = \G_0$, this amounts to
replacing the slow and fast self-energies by the real parts of their
zero-frequency values \footnote{This choice of
  $\delta \varepsilon_n^X$ aims to correctly describe low-energy
  properties, in order to solve the mass-flow problem.  More
  generally, the $\delta\varepsilon_n^X$ may be viewed as fit
  parameters that optimize the truncated CFE representation of
  $\G^\bath(\omega)$.}.  Therefore ${\rm Re} [\selfE^\bath(0)]$, the
real part of the zero-frequency self-energy of $\G^\bath$, is
reproduced correctly \footnote{For the single-impurity Anderson
    model, this guarantees that the height of the zero-temperature
    Kondo resonance at $\omega=0$, which is governed solely by
    ${\rm Re} [\selfE^\bath(0)]$, is reproduced correctly,
    irrespective of the choice of $\Lambda$.}, irrespective of the length
$N$ of the RWC used to calculate $\G^\bath$.  (Since the imaginary
parts of all self-energies are neglected, dissipative effects are not
included.)  If the original bath spectrum is symmetric, 
$\Gamma^\bathspectrum (\omega) = \Gamma^\bathspectrum (-\omega)$, as
often happens for fermionic models, then    
$\delta \varepsilon_n^{\slow/\fast} = 0$. However for an asymmetric
bath function [e.g., $\Gamma^\bathspectrum (\omega<0) = 0$, as is the
case for bosonic baths], these shifts are in general nonzero.

We will henceforth consider two types of RWCs, labeled by C1 or C2
\footnote{Chains that include neither fast nor slow shifts,   
$\delta \varepsilon_n^{\fast,\slow} = 0$, yield completely
  incorrect results, see Sec.~S-3 of Ref.~\cite{supplement}.}. A C1 chain
includes only fast shifts ($\delta \varepsilon^{\slow}_N = 0$); this
turns out to lead to results qualitatively similar to those obtained
using a SWC constructed by discretizing the original bath logarithmically, as done by Wilson, and tridiagonalizing the bath Hamiltonian $\Ham^\bath$. A C2 chain includes both the fast and
slow shifts from \Eq{eq.EnergyShift}, thus correctly reproducing
${\rm Re} [\selfE^\bath (0)]$.


\textit{Dissipative harmonic oscillator.} 
As a first example, consider a DHO with Hamiltonian
$\Ham^\imp_\DHO + \Ham^\bath$, where
\begin{equation}
  \mathcal{H}^\imp_\DHO =
\Omega a^\dagger a+ \tfrac{1}{2}(a+a^\dagger)
\bigl[
\epsilon + t_\imp (b^\pdag + b^\dagger)
 \bigr]
\end{equation}
describes an ``impurity''  oscillator with bare
frequency $\Omega$ and displacement force $\epsilon$, linearly coupled
to a bosonic bath.
The bath spectral function has the form
\begin{equation}
  \Gamma^\bathspectrum (\omega) = 2\alpha\omega_\cut^{1-s}\omega^s, \quad 0<\omega<\omega_\cut \, ,
\label{eq:boson-bath-spectrum}
\end{equation}
where $s>-1$, $\alpha$ characterizes the dissipation strength and
$\omega_{\cut} $ is a cutoff frequency, henceforth set to unity.  This
model is exactly solvable. The static impurity susceptibility at
temperature $T$, defined by 
$\chi_\stat (T) = \frac{d\langle
  a+a^\dagger\rangle_T}{d\epsilon}|_{\epsilon=0}$,
turns out to be temperature-independent and given by \cite{Vojta2010}
$\chi_\exact (T) = 1/\Omegar$, where
$\Omega_r = \Omega+{\rm Re}[\G^\bath(\omega=0)]$ can be interpreted as
the renormalized impurity frequency, reduced relative to the bare one by
the coupling to the bath. It vanishes at the critical coupling
$\alphac = s\Omega/(2\omega_\cut)$, beyond which the model becomes
unstable.

When $\chi(T)$ is computed numerically for
  $\alpha\!  < \! \alphac$ using NRG to perform thermal averages on
  SWCs of length $N_T$, one does not obtain a constant but a
  temperature-dependent curve \cite{Vojta2005,Vojta2009,Vojta2010}.
  We find the same using NRG on C1-RWCs of length $N_T$
  (Fig.~\ref{Fig.SuscCompare}, circles).  The reason is the neglect
of the TBMs associated with sites $n> N_T$:
their contribution to the renormalization shift
\mbox{${\rm Re}[\G^\bath(\omega=0)]$} in $\Omegar$ is missing.  The approach
developed above offers a straightforward cure: We simply compute
$\chi(T)$ using C2-RWCs of length $N_T$, thus incorporating the energy
shift induced by the remaining TBMs via the slow-mode shift for site
$N_T$. Since the latter substantially affects the low-energy spectrum,
these calculations require VMPS methods (see Secs.~S-2~B and S-2~C of
Ref.~\cite{supplement} for details). They yield $T$-independent $\chi$
values (Fig.~\ref{Fig.SuscCompare}, squares), in excellent agreement
with the exact ones (Fig.~\ref{Fig.SuscCompare}, solid lines).
   
We remark that SWCs constructed using previous discretization
  schemes \cite{Bulla2003,*Bulla2005,Campo2005,Zitko2009} either strongly over-
  or underestimate the critical coupling $\alphac$, reflecting the
  presence of discretization artifacts.  In contrast, our C2-RWCs
  yield $\alphac$ values that match the analytic results almost
  perfectly (see Sec~S-3~D of Ref.~\cite{supplement}).  Thus, our RWC
  construction constitutes a general new discretization scheme
 free of the discretization artifacts of previous schemes.


\textit{Spin-boson model.} Next, we consider the SBM, which is not
exactly solvable. In its Hamiltonian
$\Ham^\imp_\SBM \!+\! \Ham^\bath$,
\begin{equation}
\Ham_\SBM^\imp =
-\tfrac{1}{2} \Delta  \hat{\sigma}_x
+ \tfrac{1}{2}  \hat{\sigma}_z \bigl[
\epsilon +  t_\imp(b + b^\dagger) \bigr]
\label{hsbm}
\end{equation}
describes a spin-$\frac{1}{2}$ ``impurity'' ($\hat \sigma_i$ being Pauli
matrices) linearly coupled to a bosonic bath, with
$\Gamma^\bath(\omega)$ 
again given by \Eq{eq:boson-bath-spectrum}.  $\epsilon$ and $\Delta$
denote the bias and the tunnel splitting of the impurity spin,
respectively.

For the \textit{sub-Ohmic}
case ($0\!<\!s\!<\!1$), increasing $\alpha$ at zero temperature drives
the SBM through a quantum phase transition (QPT) from a delocalized to
a localized phase (with $\langle \hat \sigma_z \rangle_0 = 0$ or $\neq 0$,
respectively). According to a quantum-to-classical correspondence
(QCC) argument \cite{Vojta2005,Vojta2009,Vojta2012}, this QPT belongs
to the same universality class as that of a classical one-dimensional Ising chain
with long-ranged interactions \cite{Fisher1972,*Luijten1997}. Thus, the
critical exponents characterizing the QPT follow mean-field
predictions for $s \! \leqslant \! 0.5$ and obey hyperscaling for
$0.5 \! < \! s \! < \!  1$. The
QCC predictions were confirmed numerically using Monte
Carlo methods \cite{Winter2009} or sparse polynomial bases
\cite{Alvermann2009}.

\begin{figure}[t]
\includegraphics[width=0.99\linewidth]{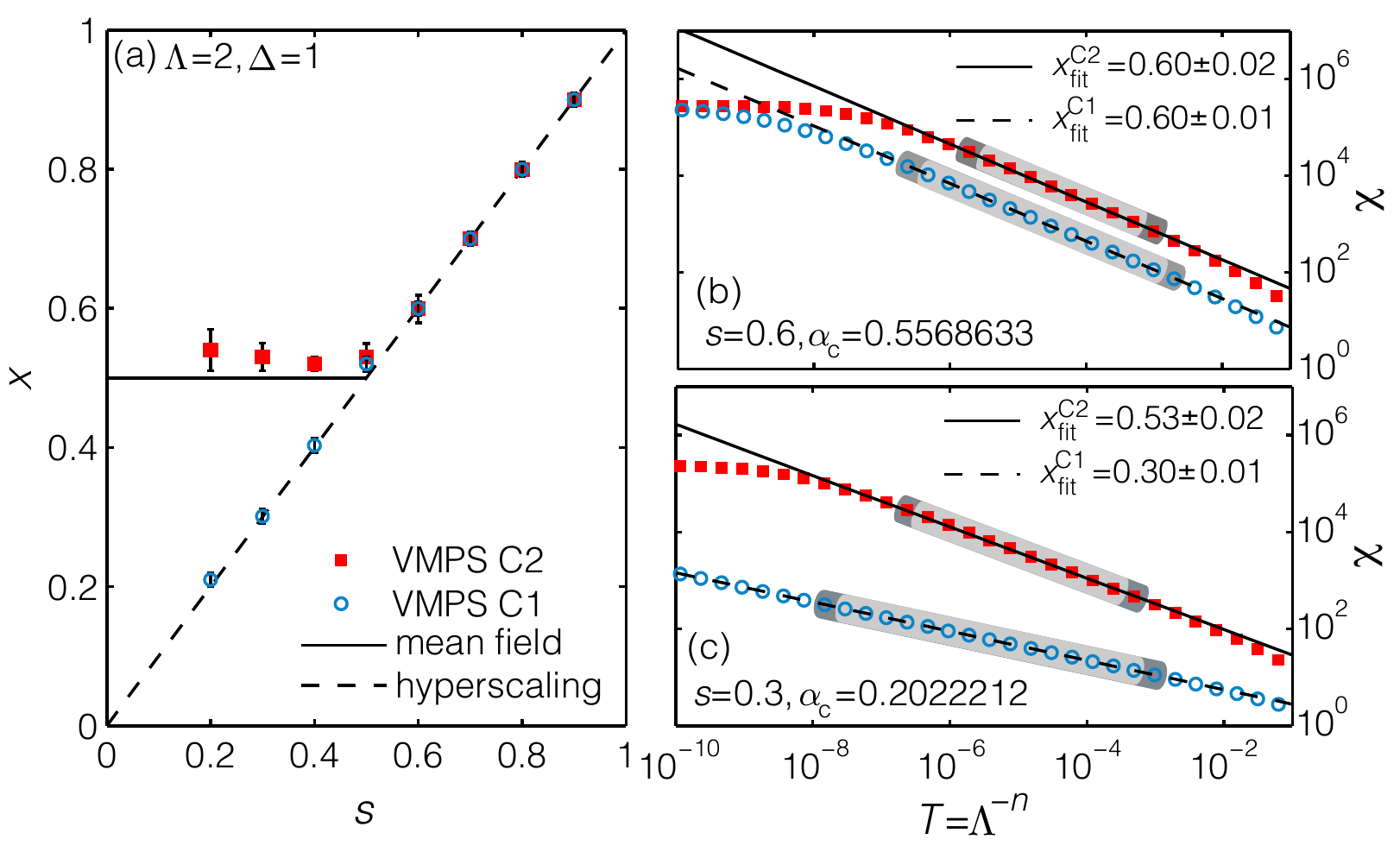}
\vspace{-7mm}
\caption{(a) Critical exponent
  $x$ 
  for the sub-Ohmic SBM, as a function of $s$, computed by VMPS using
  RWCs of type C1 (circles) and C2 (squares).  Examples of $\chi(T)$
  curves used to extract these exponents are shown in (b)
  for $s=0.3$ and (c) for $s=0.6$.  Error bars in (a) are derived by varying the fitting ranges, e.g.,~as indicated by dark and light shading in (b) and (c).
\label{fig:xSBM}} \vspace{-7mm}
\end{figure}

In contrast, verifying the QCC predictions using NRG turned out to be
challenging. Initial NRG studies \cite{Vojta2005} yielded
non-mean-field exponents for $ s \! < \! 0.5$,
but were subsequently \cite{Vojta2009,Vojta2010} found to be
unreliable, due to two inherent limitations of NRG. The first was a
too severe NRG truncation of Hilbert space in the localized phase; it
was overcome in Ref.~\cite{Guo2012,*Bruognolo2014} by using a VMPS
approach involving an optimized boson basis
\cite{Zhang1998,Weisse2000,Nishiyama1999} on a SWC, which reproduced
QCC predictions for critical exponents characterizing zero-temperature
behavior. The second NRG limitation
was the mass-flow problem: For exponents describing \textit{finite}-temperature
critical behavior at $\alpha = \alphac$, it causes
NRG on SWCs to yield hyperscaling results not
only for $0.5 \! < \! s \! < \!1 $ (correct) but also
for $s \! < \! 0.5$ (incorrect).
For example, consider the susceptibility
$\chi (T) = \frac{d\langle \hat \sigma_z
  \rangle_T}{d\epsilon}|_{\epsilon=0}$,
which scales as $\chi(T) \propto T^{-x}$ at the critical coupling
$\alphac$.
The QCC predicts $x=0.5$
for $s \! < \! 0.5$ and $x = s$ for $0.5 \! < \! s \! < \! 1$.
In contrast, past NRG calculations yielded $x=s$ throughout the interval
$0\!  <\! s \! < \! 1$
\cite{Bulla2005,Vojta2009,Vojta2010}.  We recover the latter behavior
if we compute $\chi(T)$ via VMPS calculations on
length-$N_T$ C1-RWCs [\Fig{fig:xSBM}(a), circles].  In contrast, if we
use length-$N_T$ C2-RWCs instead, the results for $x$
[\Fig{fig:xSBM}(a), squares] agree well with QCC predictions, showing
that the mass-flow problem has been cured.

\begin{figure}[t]
\includegraphics[width=0.99\linewidth]{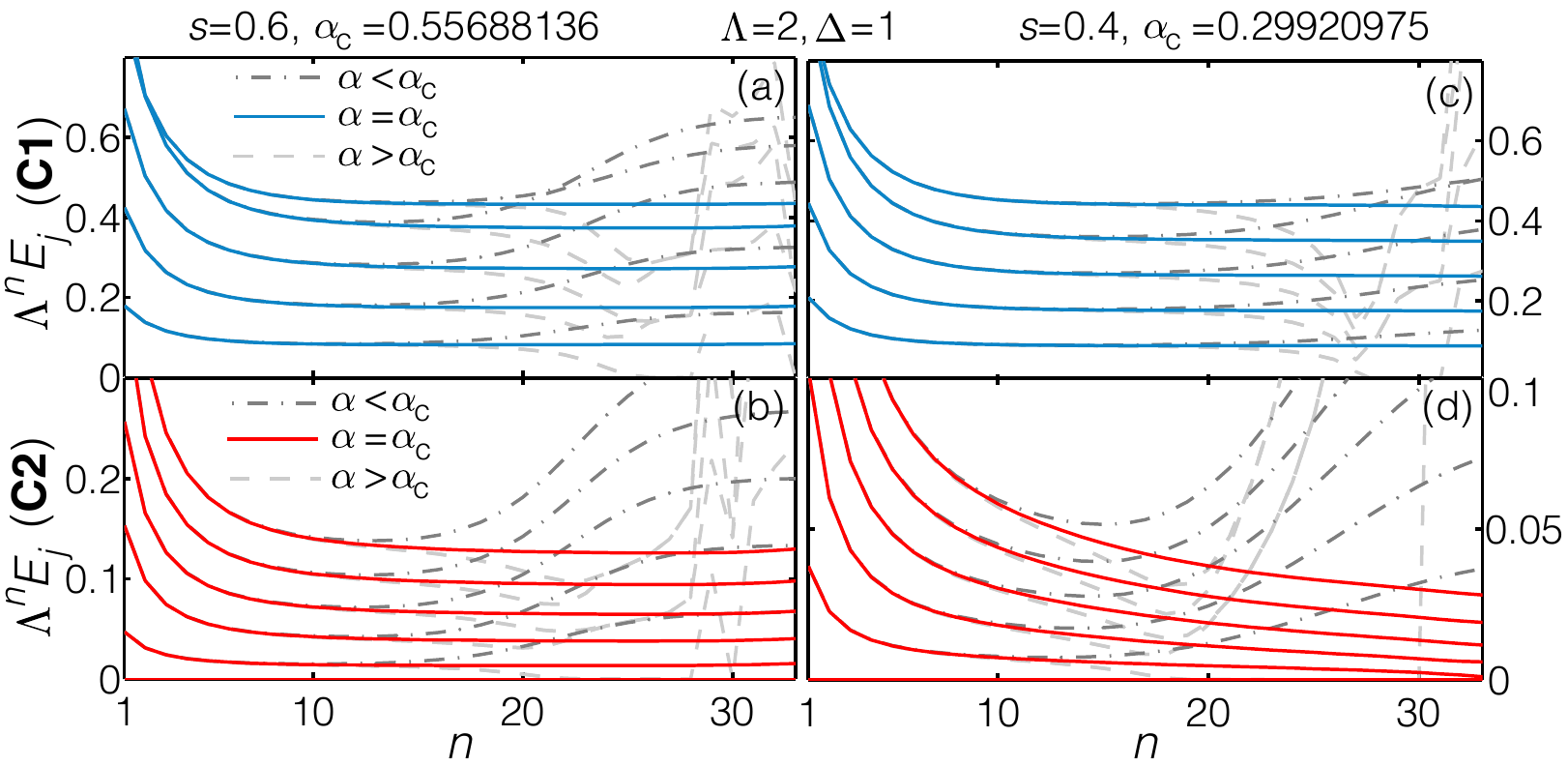}  \vspace{-6mm}
\caption{Energy-level flow diagrams for the sub-Ohmic
  SBM with $s=0.6$ (left column) and $s=0.4$ (right column), computed
  by VMPS techniques \cite{Pizorn2012,supplement}
on C1-RWCs (top row) and C2-RWCs (bottom row).
  Dashed lines depict flow to
  delocalized ($\alpha<\alphac$) or localized fixed points
  ($\alpha>\alphac$), and solid lines depict critical flow
  ($\alpha = \alphac$).
For the latter, the C2 flow in (d) is characteristic of a Gaussian fixed point.
\label{fig:Gflow}} \vspace{-6.5mm}
\end{figure}

\textit{Critical energy-level flow diagrams.} The reason why the
sub-Ohmic SBM shows qualitatively different critical behavior for
$ 0.5 \! < \! s \! < \! 1$ and $ s \! \leqslant \! 0.5$
is that the critical fixed point
is interacting for the former but Gaussian for the latter
\cite{Vojta2010}. To elucidate the difference, Fig.~\ref{fig:Gflow}
shows energy-level flow diagrams, obtained by plotting the rescaled
lowest-lying energy eigenvalues of length-$N$ Wilson chains,
$\Lambda^N E_j$, as functions of $N$.
For $s=0.6$ (left column), having an interacting critical fixed point
for which mass-flow effects are not relevant, the critical level flows
for RWCs of type C1 and C2 are qualitatively similar
[Figs.~\ref{fig:Gflow}(a) and \ref{fig:Gflow}(b)], becoming stationary (independent of $N$)
for large $N$, in a manner  familiar from fermionic NRG.

In contrast, for $s=0.4$ (right column), having a Gaussian fixed point
for which mass-flow effects do matter, the critical C1 and C2 level
flows are very different: Whereas the C1 flow becomes stationary
[Fig.~\ref{fig:Gflow}(c)] (an artifact of neglecting slow-mode
shifts), the low-lying C2 levels all flow towards zero
[Fig.~\ref{fig:Gflow}(d)], causing the level spacing to decrease
towards zero, too.
This striking behavior, inaccessible when using SWCs, is
characteristic of a Gaussian fixed point: It implies that the
fixed-point excitation spectrum contains a zero-energy bosonic
mode. Remarkably, our C2-RWCs yield a \textit{quantitatively}
correct description of the 
critical spectral flow  for $0<s<0.5$: It follows
a power law
  $\Lambda^n E_j \propto \varepsilon_n^\kappa$ with $\kappa=(2s-1)/3$,
  in perfect agreement with the prediction from controlled perturbative RG for a
  $\phi^4$-type theory with a dangerously irrelevant quartic coupling
  (see S-4~D of Ref.~\cite{supplement}).

  \textit{Conclusions and outlook.} Open Wilson chains are
  representations of quantum impurity models that achieve energy-scale
  separation while fully keeping track of the effects of bath modes,
  by iteratively replacing them by a sequence of separate baths at
  successively lower energy scales, one for each chain site. Starting from such a fully open system, the effects of these
  baths can be included systematically.
We have taken the first step in that direction, using the bath-induced
energy shift for each site to define a renormalized Wilson
chain. Remarkably, this simple scheme is sufficiently accurate to
yield renormalized impurity properties free from the long-standing
mass flow problem.  The next step, namely integrating out each
  site's bath more carefully, should lead to a description of
  dissipative effects on Wilson chains, as required for nonequilibrium
  situations. For example, the effect of bath $F_n$ on the eigenstates
  of a length-$n$ subchain could be treated using some simple
  approximation capable of mixing and broadening the eigenlevels
  (e.g.\ an equation-of-motion approach with a decoupling
  scheme). This is left for future work. 

  Finally, we note that our iterative construction of renormalized
  Wilson chains constitutes a well-controlled new discretization
  scheme that offers progress on two further fronts, unrelated to
  finite-size effects but relevant, e.g., when using NRG or DMRG as impurity
  solvers for dynamical mean-field theory
  \cite{Bulla1999,Pruschke2005,Stadler2015b,Wolf_PRX_2015}, or to study
  multi-impurity models \cite{Mitchell2015a,*Mitchell2015}. First, it
  avoids the discretization artifacts known to arise when conventional
  schemes \cite{Wilson1975,Bulla2003,*Bulla2005,Campo2005,Zitko2009} are used to
  treat strongly asymmetric bath spectra. Second, it can be
  generalized straightforwardly to treat multi-flavor models having
  nondiagonal bath spectral functions (see Sec.~S-1~B of Ref.~\cite{supplement}). \\

  We thank Andrew Mitchell for a
  stimulating discussion on discretizing multiflavor impurity models. This  
  research was supported by the DFG through the Excellence Cluster
  ``Nanosystems Initiative Munich'', SFB/TR~12, SFB~631, SFB~1143
  (M.V.), AN 275/8-1 (F.B.A), and WE4819/2-1 (A.W.).
  
  \vspace{-4mm}

\bibliographystyle{apsrev4-1}
\bibliography{biblio-OpenWilson}


\begin{center} 
\textbf{\large Supplementary material}
\end{center}
\setcounter{equation}{0}
\setcounter{figure}{0}
\renewcommand{\theequation}{S\arabic{equation}}
\renewcommand{\thefigure}{S\arabic{figure}}
\renewcommand{\thesection}{{S-\arabic{section}}} 

The supplementary material presented below deals with four
topics. Section~\ref{sec:constructing-OWC} offers a more detailed
discussion of the construction of continued-fraction expansions 
and open Wilson chains. Section~\ref{sec:numerics} describes
the numerical VMPS techniques used. Section~\ref{sec:DHO} is devoted
to a detailed study of the dissipative harmonic oscillator, in order
to benchmark our numerical methods against exact results.
Section~\ref{sec:RGflow} describes how the RG flow towards the
Gaussian fixed point of the sub-ohmic spin-boson model for
$0<s\leqslant 0.5$ can be understood using scaling arguments.

\section{Detailed discussion of CFE and OWC construction}
\label{sec:constructing-OWC}

Below we supply some technical details involved in the 
construction of (i) the continued-fraction expansion (CFE) and  
(ii) the open Wilson
chain (OWC) presented in the main text.
We begin in Subsection~\ref{sec:constructing-OWC-one-band} with
  the case of a bath involving only a single flavor of excitations, as
  discussed in the main text. In
  Subsection~\ref{sec:constructing-OWC-many-band}, we generalize the
  construction to a multi-flavor bath having a nondiagonal bath
  spectrum.

\subsection{Single-flavor bath}
\label{sec:constructing-OWC-one-band}

(i) \textit{Continued-fraction iteration step}.--- 
We here give some details on the central step of the CFE,
which takes a  retarded correlator $\G_n$
as input and produces as output a self-energy,
split into low- and high-energy contributions.  

The input correlator $\G_n$, being 
retarded, has the spectral representation
$\G_n(\omega) =  \int d \bar \omega\, \frac{\A_n(\bar
   \omega)}{\omega - \bar \omega + i 0^+}$, 
with a  spectral function,
  $\A_n(\omega) = - \frac{1}{\pi} {\rm Im} \, \G_n(\omega)$
that is  normalized to unity, 
$\int d \omega \A_n(\omega) = 1$. 
If this correlator is
  represented in the form [\Eq{eq:defineretarded}]
\begin{align} \label{eq:defineretarded_app}
\G_n(\omega) = \frac{1}{\omega - \varepsilon_n - \selfE_n(\omega)} \, ,
\end{align}
with $\selfE_n(\omega)$ analytic in the upper half-plane as required
for a retarded  self-energy, then the constant in the
denominator must be equal to the average
energy of the spectral function $\A_n(\omega)$,
$\varepsilon_n = \int d \omega \, \omega \A_n(\omega)$. 
In the main text this fact was used, but not explained.
To understand its origin, invert
\Eq{eq:defineretarded_app}, multiply it by $\G_n(\omega)$, and
integrate over frequency:
\begin{align} 
\nonumber 
 & \hspace{-2mm} 
\int \!  d \omega \,   \selfE_{n} (\omega) \, \G_{n}(\omega)  = 
 \int  d \omega \, \Bigl[( \omega - \varepsilon_n) \, 
\G_{n}(\omega) -1 \Bigr] 
\\
\nonumber 
& \qquad \qquad = 
\int \! d \omega \! \int \! d \bar \omega 
\left[ \frac{\omega - \varepsilon_n}{\omega - \bar \omega + i 0^+} - 1\right]
\A_{n}(\bar \omega)
\\ 
\nonumber 
&  \qquad \qquad = 
\int \! d \bar \omega \! \int \! d \omega 
\left[ \frac{\bar \omega - \varepsilon_n}{\omega - \bar \omega + i 0^+} \right]
\A_{n}(\bar \omega)
\\ 
&  \qquad \qquad =  
- i \pi \! \int \! d \bar \omega \, 
( \bar \omega - \varepsilon_n ) \, \A_{n}(\bar \omega)  \, . 
\label{eq:intGRGRzero}
\end{align}Since both $\G_n(\omega)$ and $\selfE_n(\omega)$ 
are  by assumption retarded functions and hence
analytic in the upper half-plane, 
the left-hand side of the first line yields zero, as can be seen by
closing the integration contour in the upper half-plane.  The second
line follows from the right-hand side of the first using the spectral
representation of $\G_{n}$, and the fact that $\A_{n}$ is normalized
to unity. Since the last line, being equal to the first, must
  equal zero too, it fixes $\varepsilon_n$ to the value stated in
  above 
(again using the unit normalization of
  $\A_n$). Once $\varepsilon_n$ has been fixed, the self-energy is
  fixed, too, by inverting \Eq{eq:defineretarded_app}:
\begin{align}
\label{eq:invert_G_to_get_Delta}
\selfE_n(\omega) = \omega - \varepsilon_n - 1/\G_n(\omega) .
\end{align}

To summarize: The fact that the retarded correlator $\G_n(\omega)$ is
analytic implies the same for its self-energy $\selfE_n(\omega)$; this
uniquely fixes $\varepsilon_n$ and thus also $\selfE_n(\omega)$
itself, which in turn can be viewed as a correlator with its own
self-energy, etc. Thus, the analyticity of $\G_n(\omega)$ guarantees
that it is always possible to iteratively construct a CFE for it. The
new twist added to this well-known fact in the present work is to zoom
in to small energies by splitting the self-energy into slow and fast
parts and using only the former as input for the next iteration step.

To explicitly implement this splitting, given 
by \Eq{eq:shortendedsplit-bath}, 
\begin{align}
\selfE_n(\omega) = 
\selfE_n^\slow (\omega) \! +\!  \selfE_n^\fast (\omega) , 
\;\; \selfE_n^X (\omega) = |t_n^X|^2 \G_n^X(\omega) ,
\phantom{.} 
\end{align}
  we proceed as follows.  We split
  $\Gamma_n(\omega) = - \frac{1}{\pi} \text{Im} \selfE_n(\omega)$,
  which may be viewed as the bath spectrum of iteration $n$, into slow
  and fast parts, 
$\Gamma_n = \Gamma_n^\slow + \Gamma_n^\fast$, with
\begin{align} 
\Gamma^X_n (\omega) = w^X_n(\omega) \Gamma_n (\omega). 
\end{align} 
Here the splitting functions $w_n^{\slow/\fast} (\omega)$ are defined
on the support of $\Gamma_n$, take values in the interval $[0,1]$, satisfy
$w_n^\slow (\omega) + w_n^\fast (\omega) = 1$, and have weight
predominantly at low/high energies. 
Then we write the split bath spectra as
$\Gamma^{X}_n (\omega) = |t_n^X|^2 \A^X_n (\omega)$, 
with ``couplings'' $t_n^X$ chosen as 
\begin{align}
|t_n^X|^2 = \int d \omega \, \Gamma^X_n(\omega) \, ,
\end{align}  
to ensure that the new spectral functions
$ \A_n^X(\omega)$ are normalized to unity. 
Using them to define
new retarded correlators via
$\G_n^X(\omega) = \! \int d \bar \omega\, \frac{\A_n^X(\bar
  \omega)}{\omega - \bar \omega + i 0^+} \, $,
we obtain the desired slow/fast splitting of the self-energy stated
above.

Next  we describe the choice of splitting functions
  $w_{n}^X (\omega)$ used to obtain 
 the numerical  results of the main text.
Let $I^\slow_n = [\omega_{\slow n}^-, \omega_{\slow n-}^+] $
denote the support of the slow spectral function
$\A_n^\slow$. The bath spectrum for iteration $n$, 
  $\Gamma_n(\omega) = -\frac{1}{\pi} \text{Im} \, \Sigma_n(\omega)$,
has support on the same interval, say $I_n$, as the
correlator $\G_n = \G_{n-1}^\slow$, i.e.\  
  $I_n = I_{n-1}^\slow$. 
To implement the splitting
$\Gamma_n = \Gamma_n^\slow + \Gamma_n^\fast$, we partition
this interval   into disjoint slow and fast subranges,
  $I_n = I_n^\slow \cup I_n^\fast$, 
with  $|\omega^\pm_{\slow n}| \leqslant |\omega^\pm_{\slow n-1}|$, and
use corresponding step-form  splitting functions:
\begin{eqnarray}
\label{eq:splittingchoice}
w_n^X (\omega) = 
\begin{cases}
1  \; \text{for} \;  \omega \in I_n^X , \\
\text{otherwise.}
\end{cases}
\end{eqnarray}

To ensure energy-scale separation, $I^\slow_n$ should
be chosen such that
 \begin{eqnarray}
  \label{eq:energy-scale-separation}
  \max\{ |\varepsilon_{n}|,|t^\slow_{n}| \} \leqslant
\max\{ |\varepsilon_{n-1}|,|t^\slow_{n-1}| \} /\Lambda \qqph
\end{eqnarray}
holds, with $\Lambda > 1$.  If the bath spectrum
$\Gamma^\bath(\omega)$ has a flat or power-law form, a natural
choice is $\omega_{\slow n}^\pm = \omega_{\slow n-1}^\pm/\Lambda$.
This is the choice used for the numerical work in the main text.
  However, if $\Gamma^\bath(\omega)$ has nontrivial structure,
  the choices for the subrange boundaries $\omega_{\slow n}^\pm$ might
  have to be fine-tuned to ensure \Eq{eq:energy-scale-separation} at
  each iteration. More generally, one might also explore using
  splitting functions $w_n^X(\omega)$ of smoother shape
  than those of \Eq{eq:splittingchoice}. The
  freedom of choice available for ensuring
  \Eq{eq:energy-scale-separation} is one of the major strengths of the
  above strategy for generating a CFE.

  (ii) \textit{Construction of open Wilson chain.}--- Here
    we provide some details on the construction of the OWC Hamiltonian
    of $\Ham^\OWC_N$ of \Eq{eq:shortendedH-OWC}.  It describes a chain
    with $N+1$ sites, each coupled to a bath of its own, and site 0
    coupled to the impurity (site $-1$) [Fig.~\ref{fig:OWC}(c)].
   It is constructed such that the free ($t_{\rm imp} =0$) correlator
  of site 0 is given by a depth-$N$ CFE,
  $\G_0 = \G^\bath$.

 We associate with each pair of
  correlators $\G_n^{\slow/\fast}$ from the CFE two mutually independent baths
  $S_n$ and $F_n$. We regard $\G^X_n$ as the free retarded correlator
  of a normalized bath operator $b_{Xn}^\dagger$, whose dynamics is
  generated by a bath Hamiltonian $\Ham_n^X$, chosen such that 
$\G^X_n(\omega) = \langle\! \langle b_{Xn}|| b_{Xn}^\dagger \rangle\!\rangle_\omega$
has the form found via the CFE.

 We start our OWC construction by associating bath $S_{-1}$ with the
original bath [Fig.~\ref{fig:OWC}(b)], setting
$\Ham_{-1}^S \! = \! \Ham^\bath$, $b^\dagger_{S,-1} \!=\! b^\dagger$
and $\G^\slow_{-1} \!=\! \G^\bath$, with impurity-bath coupling
$t_{-1}^\slow \!=\! t_\imp$. 
We then proceed iteratively, starting with $n=0$. The central
CFE iteration step of
writing $\G^\slow_{n-1}$ in the form of \Eq{eq:defineretarded}
corresponds, on the level of the Hamiltonian, to replacing the bath
$\slow_{n-1}$ by a new site $n$ [Fig.~\ref{fig:OWC}(c)], with energy
$\varepsilon_n$ and normalized site operator $f^\dag_n$, which is
linearly coupled to two new baths, $S_n$ and $F_n$, in such a way that
its free ($t_{n-1}^S \!=\! 0$) site correlator $\G_n$ equals
$\G_{n-1}^\slow$ [\Eq{eq:defineretarded}].  To achieve this, we make
the replacements $b^\dagger_{\slow n-1}\to f^\dag_n$ and
\begin{eqnarray}
  \label{eq:shortendedreplacement-for-H^slow_n}
   \Ham_{n-1}^{\slow} \to 
    \varepsilon_n f^\dagger_n   f^\pdag_n \! + \!
   \sum_X( b^\dagger_{X n} t^X_n f^\pdag_n \! + {\rm H.c.}) + \!
   \sum_X \mathcal{H}^X_n \; .  \qqph
\end{eqnarray}

 \vspace{-2mm}
\noindent
Then
$\G_n = \langle\! \langle f_n|| f_n^\dagger \rangle\!\rangle_\omega$
indeed matches \Eq{eq:defineretarded}, since the self-energy generated
for it by the new baths,
$ \Sigma_n(\omega) = \sum_X |t^X_n|^2 \G^X_n(\omega)$, agrees with
\Eq{eq:shortendedsplit-bath}.  Since $\G_n = \G^\slow_{n-1}$,
$f^\dag_n$ and $b^\dag_{\slow,n-1}$ have the same dynamics, i.e.\ the
new site, bath $S_n$ and bath $F_n$ jointly have the same effect on
site $n\! -\!  1$ as the previous bath $S_{n-1}$.  Now we iterate: we
retain the fast bath $F_n$, but replace the slow bath $S_n$ by a new
site $n+1$ coupled to new slow and fast baths $S_{n+1}$ and $F_{n+1}$,
etc. After
$N \! + \! 1$ steps, the initial $\Ham$ has been replaced by
the OWC Hamiltonian $\Ham^\OWC_N$ given in \Eq{eq:shortendedH-OWC}.

The above argument does not require 
the free Hamiltonians $\Ham^X_n$ and bath
  operators $b^\dagger_{Xn}$ to be constructed explicitly. For concreteness we specify them 
  nevertheless:
\begin{eqnarray}
  \label{eq:freeHn-fn}
  \mathcal{H}_n^X = \sum_q \varepsilon^X_{qn} b^\dagger_{Xqn} b^\pdag_{Xqn} , 
 \quad
 b^\dagger_{Xn} = \sum_q
  b_{Xqn}^\dagger  \lambda^{X}_{qn}\; . \quad  
\end{eqnarray}
These involve a set of canonical annihilation and creation operators
  satisfying $[b^\pdag_{Xqn}, b^\dagger_{Xqn}]_{\pm} = 1$ ($+$ for a
  fermionic anti-commutator, $-$ for a bosonic commutator).  
The bath operators $b^\pdag_{Xn}$ are normalized to 
satisfy $[b^\pdag_{Xn}, b^\dagger_{Xn}]_{\pm} = 1$. 
The free dynamics of $b^\dagger_{Xn }$, generated by
$\Ham_n^X$, is characterized by the free retarded correlator
and spectral function
\begin{subequations}
\begin{align} 
\label{eq:G0bath} 
\G_n^X(\omega) & =
\sum_q
\frac{|\lambda^X_{qn}|^2 }{\omega - \varepsilon^X_{qn} + i 0^+} ,
\\
\A^X_n(\omega) & = \! \sum_q
 \! |\lambda^X_{qn}|^2 \delta(\omega - \varepsilon^X_{qn}) \, . 
\end{align}
\end{subequations}
The bath energies $\varepsilon^X_{qn}$ and
couplings $\lambda^X_{qn}$ are assumed such that 
$\G_n^X(\omega)$ has the form obtained in the CFE.

\subsection{Multi-flavor bath}
\label{sec:constructing-OWC-many-band} 

Next we consider impurity models involving a multi-flavor bath with
$\mflavor$ flavors of excitations, labeled by an index
$\nu = 1, \dots, \mflavor$. We assume that the impurity
  Hamiltonian $\Ham_\imp[ b_\nu^\dagger]$, describing the impurity
  degrees of freedom and their coupling to the bath, depends on the
  bath only through $\mflavor$ bath operators $ b^\dagger_{\nu}$ and
  their conjugates $ b_{\nu}$, not necessarily normalized or
  orthogonal, with retarded correlator
  $\G^\bath_{\nu \nu'} (\omega) = \langle \! \langle b^\pdag_{\nu};
  b_{\nu'}^\dagger \rangle \!  \rangle_\omega$.
  We assume that the corresponding bath spectrum,
\begin{eqnarray} \Gamma^\bath_{\nu \nu'} (\omega) = -
    \bigl[\G_{\nu \nu'}^\bath(\omega) - \G^{\bath \ast}_{\nu' \nu}
    (\omega)\bigr] / (2 \pi i ) \, ,      
 \end{eqnarray} 
 is a specified, Hermitian, positive definite \textit{matrix} function
 (i.e.\ for any given $\omega$, the eigenvalues of the matrix are real
 and non-negative).  Together with the form of $\Ham_\imp$, this
 matrix function fully determines the impurity dynamics. Models
   of this structure arise in studies of the Kondo compensation cloud
 \cite{Borda2007,*Lechtenberg2014}, when considering
   multi-impurity situations \cite{Mitchell2015a,Mitchell2015}, and
 in DMFT studies of multi-band lattice models, where
 $\Gamma^\bath_{\nu \nu'}(\omega)$ is constructed iteratively from the
 impurity spectral function $\A^\imp_{\nu \nu'}(\omega)$ computed at
 the previous DMFT iteration.   
    
    If $\Gamma^\bath_{\nu \nu'}(\omega)$ can be diagonalized
      using a frequency-independent unitary transformation, the
      eigenvalues, say $\Gamma_{\nu}(\omega)$, constitute $\mflavor$
      hybridization functions that can be discretized independently,
      using either standard Wilsonian discretization or our RWC
      discretization scheme. Here we are interested in the more
      general case that diagonalizing the bath spectrum requires a
      frequency-dependent unitary transformation,
      $\Gamma^\bath_{\nu \nu'} (\omega)= \sum_{\bnu}
      u^\dagger_{\nu \bnu} (\omega) \Gamma_{\bnu} (\omega) u_{\bnu
        \nu'} (\omega)$.
    This would be the case, for example, for DMFT studies of a
    fermionic lattice model with broken band degeneracy and spin-orbit
    coupling; the corresponding self-consistent impurity model is a
    multi-band Anderson model involving nondiagonal level-bath
    couplings, leading to a nondiagonal impurity spectral function.

    To treat this situation in Wilsonian fashion, 
    one could write the bath
    spectrum as
    $\Gamma^\bath_{\nu \nu'} (\omega) = \int d \varepsilon_q
    \sum_{\bnu} v^\dagger_{q \nu \bnu} \, \delta(\omega -
    \varepsilon_q) \, v_{q \bnu \nu}$,
    with bath-lead matrix elements
    $v_{q \nu\nu'} = \sqrt{\Gamma_{\nu}(\varepsilon_q)} \, u_{\nu
      \nu'} (\varepsilon_q)$,
    and discretize the integral logarithmically (with the implicit
    assumption that $\Gamma_\nu(\omega)$ and $v_{q \nu \nu'}(\omega)$
    change sufficiently slowly with $\omega$ that within a
    discretization interval they may be replaced by constants).  We
    note, though, that the neglect of truncated bath modes is
    potentially more problematic for multi- than single-flavor models,
    since $\Gamma_{\nu \nu'}^\bath(\omega)$ will generically have
    matrix elements asymmetric in frequency.

Below we explain how multi-flavor models can alternatively be
discretized using a generalization of our RWC construction. (We thank
Andrew Mitchell for a stimulating discussion which led to this
realization.) The overall strategy is completely analogous to the
single-flavor case, but with a flavor index added to all creation and
annihilation operators (e.g.\ $b^\dagger_{X n \nu}$), and two flavor indices
to all matrix elements (e.g.\ $t_{ n \nu \nu'}^X$) and correlators
(e.g.\ $\G_{n \nu\nu'}^X$). We will mostly use a compact notation that
suppresses these indices and indicates their implicit presence by an
underscore, e.g.\ $\ub^\dagger_{Xn}$, $\ut_n^X$, $\uG_n^X$,
  $(\ub^\dagger_{Xn} \ut^X_n)_{\nu'} = b^\dagger_{Xn\nu} t^X_{n\nu \nu'}$, and
  $ \uf^\dagger_n \uvarepsilon_n \uf^\pdag_n = \sum_{\nu \nu' }
  f^\dagger_{n \nu} \varepsilon_{n\nu \nu'} f^\pdag_{n\nu'} $, etc. \vspace{1mm}

  \textit{Extracting normalized modes from bath spectrum.}---
    The CFE to be constructed below involves a sequence of bath
    spectra with matrix structure, generically denoted by
    $\uGamma (\omega)$.  Each is a Hermitian, positive definite matrix
    function, $\uGamma (\omega)= \uGamma^\dagger (\omega)$.  We would
    like to express such a function in terms of a Hermitian, positive
    definite matrix function $\uA(\omega)$ that is normalized as
\begin{align}
  \label{eq:normalization-energy-multiflavor}
 \int \! d \omega \,  \uA (\omega)  = \uone \;  , 
\end{align}
because such an $\uA(\omega)$ can be viewed as the spectral function
of a set of \textit{orthonormal} bath modes. To this end, we note that
the frequency integral $\uw = \int d \omega \, \uGamma (\omega)$
yields a Hermitian, positive definite matrix. (Reason: If two
matrices are Hermitian and positive definite, the same is true for
their sum, and similarly for an integral of such matrix functions.)
The  matrix $\uw$ can thus be diagonalized in the form
$\uw = \uu^\dagger \ud \, \uu$, with $\uu$ unitary and $\ud$
diagonal and positive.  Then the matrix
$\ut = \uu^\dagger \sqrt{\ud} \, \uu$
can be used to write the bath spectrum 
in the form
\begin{align}
\label{eq:canonicalform}
  \uGamma (\omega) = \ut^\dagger  \uA (\omega)  \, \ut  \, , 
\end{align}
where both $\ut$ and  
$\uA$ are Hermitian and positive definite, while $\uA$ 
by construction is normalized as in \Eq{eq:normalization-energy-multiflavor}. 
The first moment of $\A$ yields a Hermitian matrix, too:
$ \uvarepsilon = \int \! d \omega \,  \omega \, \uA (\omega)$.
In the chain to be constructed below, $\uvarepsilon$ plays
the role of an onsite Hamiltonian and $\ut$ that of a nearest-neighbor
coupling.   If desired, one may make another unitary
transformation that diagonalizes  either $\uvarepsilon$ or
$\ut$, while leaving the normalization condition 
\eqref{eq:normalization-energy-multiflavor} in tact.

\textit{Continued-fraction expansion.---}  
As for the one-band case, we aim to iteratively
represent $\uG^\bath(\omega)$
in terms of a sequence of continued-fraction expansions 
that zoom in on low energies. 
These involve a sequence of Hermitian, positive definite functions,
$\uA_n^{X}(\omega) = \uA_n^{X \dagger}(\omega)$.
Each is normalized to unity [\Eq{eq:normalization-energy-multiflavor}]
and can be viewed as the spectral function
of a retarded correlator $\uG^{X}_n(\omega)$,
\begin{align}
\uA_{n}^X (\omega) = - 
\bigl[\uG_n^{X}(\omega) - \uG_n^{X \dagger} (\omega)\bigr]
/ (2 \pi i ) \, , 
\end{align}
which in turn can be expressed as
\begin{align} 
\label{eq:defineretarded_app_multi} \uG_n^X(\omega) =
  \int d \bar \omega \frac{\uA_n^X(\bar \omega)}{\omega - \bar \omega
    + i 0^+}  \, . 
\end{align}

The multi-band CFE construction follows the one-band case, except
that all correlators carry underscores to indicate their matrix
structure. First we initialize the CFE by expressing the bath
  spectrum in terms of a normalized spectral function,
  $\uGamma^\bath(\omega) = \ut_\imp^{\slow \dagger} \, \uA^\slow_{-1}
  (\omega) \, \ut^\slow_\imp$
  [cf.\ \eqref{eq:canonicalform}] and compute the corresponding
  retarded correlator $\uG^\slow_{-1}$ via
  \Eq{eq:defineretarded_app_multi}. Starting with iteration $n=0$, we
  then iteratively use $\uG^\slow_{n-1} $ as input to define a new
  retarded correlator $\uG_n$ and its retarded self-energy
  $\uselfE_n$,
\begin{align}
\label{eq:defineretarded-multi} 
\uG_n(\omega) = \G^\slow_{n-1}(\omega) = 
1/\left[\omega \uone - \uvarepsilon_n
- \uselfE_n(\omega) \right] \, , 
\end{align}
with $\uvarepsilon_n = \int \! d \omega \, \omega \, \uA_n(\omega)$.
Then we split this self-energy into low- and high-energy parts
by writing it as 
\begin{eqnarray}
  \label{eq:shortendedsplit-bath-appendix}
\uselfE_n = 
\uselfE_n^\slow + \uselfE_n^\fast  , 
\quad \uselfE_n^X (\omega) = \ut_n^{X\dagger} \uG_n^X(\omega) \ut_n^X .\quad  
\phantom{.} 
\end{eqnarray}
To be concrete, we achieve this splitting
by proceeding as follows. We split
$\uGamma_n(\omega) = -\big[ \uselfE_n(\omega) - \uselfE^\dagger_n
\big]/(2\pi i)$,
the bath spectrum of iteration $n$, into slow and fast parts,
$\uGamma_n = \uGamma_n^\slow + \uGamma_n^\fast$, with
\begin{align}
\label{eq:split-coupling-functions-multiflavor}
  \Gamma^X_{n, \nu \nu'} (\omega) =  w_{n, \nu \nu'}^X (\omega) 
  \Gamma_{n, \nu \nu'}(\omega)
\end{align}
(no index summation implied here), using symmetric, real
matrix functions $\uw_n^X (\omega)$. Their matrix
elements $ w_{n \nu \nu'}^{\slow/\fast} (\omega)$ are splitting
functions that are defined on the support of $\uGamma_n$,
take values in $[0,1]$, have weight predominantly at low/high
energies, and satisfy
$w^\slow_{n\nu\nu'} (\omega) + w_{n\nu\nu'}^\fast (\omega) = 1$.  
(The simplest choice would be
$\uw_n^{X}\!(\omega) = w_n^{X} \! (\omega) \uone$, using the same pair
of weighting functions for all matrix elements; but situations may
arise where the additional freedom of making different choices for
different matrix elements is useful.)
Since the splitting functions are symmetric and non-negative, the split
spectra $\uGamma_n^X$ are Hermitian and positive definite matrix
functions, too. We can thus express them in terms of normalized
spectral functions [\Eq{eq:canonicalform}]:
\begin{align}
\label{eq:split-coupling-functions-multiflavor}
 \uGamma^{X}_n (\omega) 
=  \ut_n^{X\dagger} \uA^X_n (\omega)  \ut_n^{X} .
\end{align}
Computing the  corresponding retarded correlators $\uG_n^X$
[\Eq{eq:defineretarded_app_multi}] we obtain the 
self-energy splitting stated in \Eq{eq:shortendedsplit-bath-appendix}.
To ensure energy-scale separation, the weighting
functions $\uw_n^X$ should
be chosen such that
 \begin{eqnarray}
  \label{eq:energy-scale-separation-matrix} 
  \max\{ \parallel \!\uvarepsilon_{n}\! \parallel ,
 \parallel \!\ut^\slow_{n}\! \parallel  \} \leqslant
\max\{  \parallel \!\uvarepsilon_{n-1}\! \! \parallel ,
\parallel \!\ut^\slow_{n-1}\!\!  \parallel \} /\Lambda \qqph
\end{eqnarray}
holds, with $\Lambda > 1$, where $ \parallel \;\;  \parallel $ denotes
some matrix norm.

Iterating this procedure yields a sequence of CFEs for $\uG^\bath$,
in the same fashion as for the one-band case.

\textit{Chain representation.---} The CFE data
($\uvarepsilon_n, \ut_n^X, \uG_n^X$) can now be used to represent the
model in terms of a chain with $N+1$ sites, each coupled to a bath of
its own. The chain is constructed such that the free
($\ut_{-1}^\slow=0$) correlator of the first site ($n=0$) is
given by a dept-$N$ CFE. To this end, we associate each pair of
correlators $\uG^{\slow/\fast}_n$ with two mutually independent baths
$S_n$ and $F_n$, and regard  each $\uG^X_n$ as the free retarded
correlator of a set of $m_f$  normalized bath operators $\ub^\dagger_{Xn}$, 
whose free dynamics is generated by a bath Hamiltonian
$\Ham^X_n$, such that $\uG^X_n(\omega) =
 \langle \! \langle \ub_{Xn}^\pdag|\ub^\dagger_{Xn} \rangle \!
\rangle_\omega$.
 These  free bath Hamiltonians and bath operators
have the form
  \begin{align}
    \mathcal{H}_n^X & = \! \sum_q  \ub^\dagger_{Xqn} \uvarepsilon^X_{qn} \ub^\pdag_{Xqn} \, , \quad 
  \label{eq:freeHn-fn-multiflavor}
  \ub^\dagger_{Xn}   = 
 \sum_q  \ub_{Xqn}^\dagger \ulambda^{X}_{qn}  \, ,
\end{align}   
where $\uvarepsilon^X_{qn}$ and $\ulambda_{qn}^X$ are matrices w.r.t.\
to the flavor indices. $\uvarepsilon^X_{qn}$ is diagonal and real, and
$\ulambda_{qn}^X$ unitary, normalized such that
$[\ub^\pdag_{Xn}, \ub^\dagger_{Xn}]_{\pm} = \uone$.  
The free bath
correlators and spectral functions then have the explicit representations
\begin{subequations}
\begin{align}
\label{eq:G0bath-multiflavor}
  \uG_n^{X}(\omega) & =
                      \sum_q \ulambda^{X\dagger}_{qn} 
                      \bigl[(\omega + i 0^+) \uone - \uvarepsilon^X_{qn} \bigr]^{-1} 
                      \ulambda^{X}_{qn} \, , 
  \\
  \uA_n^{X}(\omega) & =   
\sum_{q} \ulambda^{X\dagger}_{qn} \delta(\omega \uone - \uvarepsilon^X_{qn})
\ulambda^{X}_{qn} \, . 
\end{align}
\end{subequations}
This representation for $\uA_n^X(\omega)$ shows explicitly
that it is a Hermitian, positive definite matrix function.

The iterative OWC construction proceeds as for the single-flavor case,
except that all operators, matrix elements and correlators now carry
underscores to indicate implicit flavor indices. For example, the
  generalization of \Eq{eq:shortendedreplacement-for-H^slow_n} now
  involves the replacements $\ub^\dagger_{\slow n-1}\to \uf^\dag_n$
  and
\begin{eqnarray}
  \label{eq:shortendedreplacement-for-H^slow_n-multiflavor}
   \Ham_{n-1}^{\slow}  \! \to \!
   \uf^\dagger_n  \uvarepsilon_n   \uf^\pdag_n \! + \!
   \sum_X(\ub^\dagger_{X n} \ut^X_n  \uf^\pdag_n \! + {\rm H.c.}) +
   \! \sum_X \mathcal{H}^X_n  . \; \qqph
\end{eqnarray}
The final OWC Hamiltonian has the same form as \Eq{eq:shortendedH-OWC}
of the main text, suitably decorated with underscores, and with
$\Ham^\imp[\uf_0^\dagger \, \ut_\imp]$ as impurity Hamiltonian.
Similarly, when moving on to a RWC, the energy shift equation
(\ref{eq.EnergyShift}) of the main text is decorated by underscores,
i.e.\ we shift the onsite energy matrices $\uvarepsilon_n$ by
$\delta \uvarepsilon_n^X$ shifts that should be chosen to
  optimize the truncated CFE representation of $\uG^\bath$.  We
expect this step to be more important for multi- than single-flavor
models, since $\uGamma_n^X(\omega)$ will generically have matrix
elements asymmetric in frequency. If one is interested mainly
in correctly reproducing low-energy properties, one could
choose $\delta \uvarepsilon_n^X = \text{Re} \uSigma_n^X(\omega=0)$,
as in the main text. Another option would be to view
the $\delta \uvarepsilon_n^X$ as fitting parameters, chosen
to get the best possible agreement between the 
depth-$n$ CFE for $\uG^\bath(\omega)$ and its actual form.

\section{Numerical details} 
\label{sec:numerics}

In this section, we elaborate on the details of the numerical methods
employed in the main text. In Subsection~\ref{sec:bNRG} we briefly
review NRG and its limitations in the context of bosonic impurity
models. In Subsection~\ref{sec:VMPS} we discuss the VMPS techniques by
which these limitations can be overcome.  Finally, in
Subsection~\ref{sec:VMPSm} we present a generalized VMPS scheme that
simultaneously targets multiple low-energy states on the Wilson chain,
which enables us to generate the well-controlled energy-level flow
diagrams for the sub-Ohmic spin-boson model (SBM)
shown in Fig.~\ref{fig:Gflow}  of the main text.

\subsection{Bosonic NRG} \label{sec:bNRG} The numerical
renormalization group (NRG) is one of the most powerful tools to
numerically evaluate the properties of quantum impurity models
\cite{Wilson1975}. Wilson's formulation of ``standard NRG'' involves
two steps. First, the model is represented in terms of a Wilson chain,
i.e.\ a semi-infinite tight-binding chain whose hopping matrix
elements $t_n$ decrease exponentially with $n$, ensuring energy-scale
separation along the chain. In the main text and
Sec.~\ref{sec:constructing-OWC}, we have described in detail how this
is achieved for an RWC; for details on setting up a SWC we refer to
\Refs{Bulla2005} and \cite{Bulla2008}. Second, the chain is
diagonalized iteratively one site at a time, discarding high-energy
states at each step, to yield a set of so-called Wilson shells, where
shell $N$ contains the low-lying eigenstates of a finite chain whose
last site is labelled $N$ (a ``length-$N$'' chain). These shells can
be used to calculate both thermodynamic and dynamical quantities; in
particular, we employed the full-density-matrix NRG scheme (fdm-NRG)
to evaluate thermal averages of observables in this work
\cite{Weichselbaum2007}.

Whereas NRG has been highly successful in the context of fermionic
impurity models, its application to bosonic baths has been impeded by
two numerical issues, (i) the mass-flow error and (ii) the local
Hilbert space truncation. We elaborated on (i) in detail in the main
text. We add that NRG cannot be completely cured from the mass flow
using the C2-RWC construction, as discussed in more detail in
\Sec{sec:DHO} below. This is related to the iterative nature of the
NRG diagonalization procedure, which does not allow to incorporate any
feedback of the slow-mode correction to earlier iterations, in
contrast to the variational setup presented in \Sec{sec:VMPS}
below. Problem (ii) is related to the fact that only a limited number
of bosons can be included in an NRG calculation. NRG requires an
\textit{a priori} truncation of the infinite-dimensional local bosonic
Hilbert space on each site $n$ to a numerically feasible number of
$d_n$ bosonic states. For example, for the spin-boson model NRG is therefore not able to 
accurately deal with the fact that the oscillator displacement occurring in the localized
phase grows exponentially along the Wilson
chain, which implies that the number of bosons in the standard
oscillator representation must increase exponentially, too
\cite{Vojta2009}.

In the context of the sub-Ohmic SBM, it has been thoroughly illustrated  how the limitations of bosonic NRG can tamper with physical properties. Here, the interplay of these two numerical
issues affected a number of critical exponents, causing them to follow
hyperscaling instead of mean-field results for $0<s<0.5$
\cite{Vojta2005}. The internal consistency of these NRG results (which
were later shown to be incorrect) was so striking that it initially
lead to the controversial conclusion, that the quantum-to-classical
correspondence breaks down in case of the sub-Ohmic SBM. This subtle
``conspiracy of errors'' \cite{Vojta2012} implies that NRG is not
fully equipped to deal with bosonic baths, since parts of the phase
diagram and, in particular, the impurity quantum phase transition, may
not be reliably accessible for the method.

\subsection{VMPS with optimal boson basis} \label{sec:VMPS} The
intrinsic flaws of bosonic NRG can be completely dealt with by
employing the strategy of the density matrix renormalization group
(DMRG) to treat RWC Hamiltonians
\cite{White1992,*White1993,*Schollwoeck2005,Weichselbaum2009}. To this end, we use
the matrix-product-state (MPS) formulation of DMRG
\cite{Schollwoeck2011}, which we refer to as variational
matrix-product-state approach (VMPS) in the following
\cite{Weichselbaum2009,Saberi2008}. This method can overcome the issue
of Hilbert space truncation by using a flexible, shifted optimized
boson basis (OBB) \cite{Zhang1998}, as shown in
\cite{Guo2012,*Bruognolo2014,Brockt_PRB_2015,*Schroeder_PRB_2016,*Dorfner_PRA_2016}. Moreover,
the mass-flow problem can be successfully cured by performing the
variational procedure on C2-RWCs, as demonstrated in the main text. We
briefly elaborate on the main aspects of the VMPS approach and refer
to \cite{Guo2012,*Bruognolo2014} for technical details.

The goal of the VMPS approach is to efficiently represent the ground state of a Wilson chain with $N$ bath sites in the formalism of matrix-product states \cite{Schollwoeck2011}. A generic MPS of a bosonic impurity model has the form
\begin{equation}
\label{eq:MPS1}
|\psi\rangle = \sum_{\sigma,\mathbf{m}} A^{[\sigma]} A^{[m_0]} A^{[m_1]} \,... \, A^{[m_{N}]} |\sigma\rangle | \mathbf{m} \rangle\,,
\end{equation}
where $|\sigma\rangle$ represents the local space of the impurity
(e.g., a spin-$\frac{1}{2}$ degree of freedom) and
$ \mathbf{m} = |m_0\rangle ... |m_N\rangle$ describes the local boson
number eigenstates in a truncated Fock basis, i.e.,
$f^\dagger_n f^\pdag_n |\mathbf{m} \rangle = m_n |\mathbf{m} \rangle$ with
$m_n = 0,1, ... , d_n-1$. Starting with a random MPS, the ground state
is approximated by iteratively varying the tensors $A^{[...]}$ to
minimize the energy of the Wilson chain Hamiltonian, sweeping back and
forth through the chain until a global energy minimum is reached with
sufficient
convergence. 

One key advantage of VMPS over NRG is the ability to flexibly adapt
the local bosonic state basis on each site of the Wilson chain during
the optimization process. This concept of an adaptive boson basis
enables us, for example, to determine the ground state also in the
localized phase of the SBM faithfully, which is not possible in NRG
calculations. Our OBB implementation includes two features: First, we
introduce an additional basis transformation $V$ with
$V^{\dagger}V=\id$, which maps the local harmonic oscillator basis
$|m_n\rangle$ onto a smaller effective basis $|\tilde{m}_n\rangle$ on
each site $n$,
\begin{equation}
|\tilde{n}_n\rangle = \sum_{m_n=0}^{d_n-1} V_{\tilde{m}_n,m_n} |m_n\rangle \quad (\tilde{m}_n =0,\,...\,,\tilde{d}_n-1)\,.
\end{equation}
$V$ can be naturally embedded in the MPS structure and is optimized in an additional local update to determine the best set of local basis states $|\tilde{m}_n\rangle$ for the subsequent update steps \cite{Guo2012,*Bruognolo2014}.

Second, we explicitly incorporate any oscillator displacements
occurring in strong-coupling phases when constructing the local boson
basis sets. To this end, we shift the oscillator coordinate
$\hat{x}_{n}=\frac{1}{\sqrt{2}}(f_{n}+f_{n}^{\dagger})$ on
each site $n$ by its equilibrium value $\langle\hat{x}_{n}\rangle$
\cite{Alvermann2009} employing an unitary transformation to the
Hamiltonian of the system
\cite{Guo2012,*Bruognolo2014}. $\langle\hat{x}_{n}\rangle$ can be
determined self-consistently in a variational setting. Using such a
setup, the OBB is able to capture quantum fluctuations around the
shifted coordinate
$\hat{x}_{n}'=\hat{x}_{n}'-\langle\hat{x}_{n}\rangle$.

In practice, the shifted OBB not only allows a significant increase of
the size of the local basis sets from $d_n \approx \mathcal{O}(10^2)$
to $d_n \lesssim \mathcal{O}(10^{4})$ by means of the basis
transformation $V$. In addition, the shifted oscillator basis enables
us to account for the exponentially growing oscillator displacements
in a numerically quasi-exact way, which would require a local
dimension of up to $d^{\rm eff}_n \approx \mathcal(10^{10})$ in a
nonshifted basis \cite{Guo2012,*Bruognolo2014}.

An additional advantage of the variational optimization over NRG
  is the fact that the former typically involves multiple sweeps along
  the chain, so that information from different parts of the Wilson
  chain (i.e., from different energy scales) is incorporated during
  the optimization process.  This feedback mechanism
is not needed for chains that have energy-scale
  separation. However, the latter is violated 
at the last site of a C2-RWC, where the slow-mode energy shift
  is large enough to
  affect the nature of the MPS not only on the last site but also on
  several preceding sites. In contrast to NRG, the VMPS approach is
capable of feeding back this slow-mode information from low-energy
scales to higher ones during the optimization sweeps, which is key to
successfully avoid any mass-flow effects.

Even though the VMPS scheme described above only targets the
  ground state, it can be used to mimic finite-temperature averages on
  the Wilson chain, such as the thermal average
  $\langle a+a^\dagger \rangle_T$ or $\langle \hat \sigma_z \rangle_T$
  needed to compute the local susceptibility $\chi(T)$ for the DHO or
  SBM, respectively.  To this end, we compute the ground-state
  expectation value $\langle \GG | a + a^\dagger | \GG \rangle_{N_T}$
  or $\langle \GG | \hat \sigma_z| \GG \rangle_{N_T}$
  for a length-$N_T$ C2-RWC, where $N_T$ is chosen such that the
  chain's lowest energy scale matches the temperature,
  $T \sim \Lambda^{-N_T}$.  This works because, for a length-$N_T$
  chain, the response of the ground state is calculated for a discrete
  spectrum whose low-energy excitations have characteristic spacing
  $T$.  This is the strategy that was used for the VMPS calculations
  of $\chi(T)$ reported in the main text. A more detailed description
  of this strategy is given in \Sec{sec:DHO} below, devoted to a
  detailed study of the dissipative harmonic oscillator (DHO). There
  we compare several different strategies for computing thermal
  averages and benchmark their results against the exact solution for
  $\chi(T)$.

An important prerequisite for studying critical properties is a
  highly accurate determination of the critical coupling
  $\alphac$. Numerically, it can be found in several ways. First, by
  determining the $\alpha$-value at which the susceptibility
  $\chi(T=0)$ diverges; this was our method of choice in the context
  of the DHO. Second, by monitoring how the NRG or VMPS energy flow
  diagrams evolve with $\alpha$. For the SBM there exist a third
  option, namely monitoring the behavior of the average boson
  occupation per site, $\langle m_n \rangle$: at the phase boundary it
  stays almost constant throughout the chain, but in the delocalized
  (localized) phase it decreases (increases) towards the end of the
  Wilson chain. We used the third scheme for the SBM, since it can be
  automated very easily. C2-chains sometimes required additional
  fine-tuning, since the slow-mode shift always increases the
  occupation numbers at the end of the chain.

We end this section with some technical notes. All VMPS
ground-state calculations in this work for both DHO and SBM were
performed using a 1-site update with fixed bond dimension $D=60$,
$d_n=100$, and $\tilde{d}_n = 16$.  Convergence was assumed if the
change in the chain's  ground-state energy dropped below the
threshold $|\delta E_\GG| < 10^{-15}$, which for our
longest chains corresponded to $\simeq 0.5$ 
of the hopping matrix element $t_N$ to the last site.
This typically took 10 to 50 sweeps. For the
determination of the temperature-dependent susceptibility $\chi(T)$,
we performed separate VMPS calculations for each value of $T$ and used
a five-point stencil to evaluate the numerical derivative with respect
to $\epsilon$. The convergence of the results with respect to all
important numerical parameters was checked thoroughly.

\begin{figure}[t]
  \centering
\includegraphics[width=.99\linewidth]{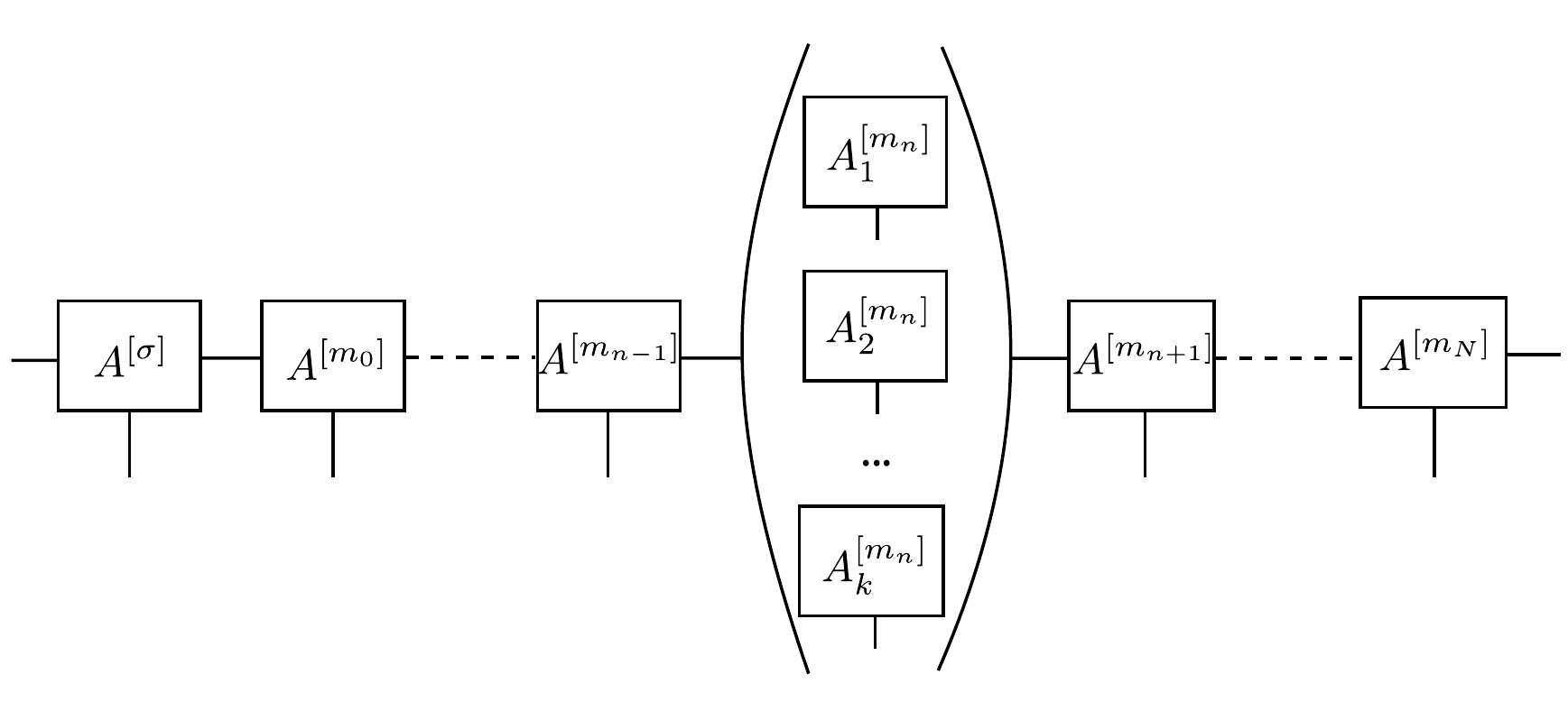} \vspace{-6mm}
\caption{Illustration of mVMPS setup for variationally calculating the $m$ lowest-energy excitation on a Wilson chain.}
  \label{Fig.mVMPS}
\end{figure}

\subsection{Multilevel VMPS} \label{sec:VMPSm} The study of
energy-level flow during the renormalization procedure is an important
part of the NRG toolbox to characterize the fixed-point properties of
an impurity model. However, in the presence of the mass-flow error,
prominent for a bosonic bath with asymmetric bath spectrum, NRG
does not correctly capture the physics of the critical fixed point and
the resulting RG flow can no longer be considered reliable. On the
other hand, we have already demonstrated that VMPS techniques are able
to appropriately deal with mass flow; below we show that they can also
be employed to properly access the energy-level flow at quantum
critical points.

In its standard formulation, described above, VMPS only targets the
ground state and does not have sufficient information about low-lying
excited states on the Wilson chain to accurately describe the
energy-level flow. In order to go beyond ground-state physics and
properly capture the critical energy-level RG flow of multiple
low-lying levels, we have implemented a multi-level VMPS (mVMPS)
optimization scheme, in the spirit of \Ref{Pizorn2012}, that
simultaneously targets the lowest $k$ energy eigenstates
$|\psi_j\rangle$. A detailed description of our procedure
may be found in Sec. 2.3.6 of \cite{Linden2014}. Here we just
outline the main idea.

Assuming canonical form of the MPS with the center shifted to site $n$,
we define an array $\mathbf{A}^{[m_n]}$ consisting of $k$
tensors $\{ A_1^{[m_n]}, A_2^{[m_n]},\,...\, , A_k^{[m_n]} \}$
(illustrated in \Fig{Fig.mVMPS}). For each tensor $A_j^{[m_n]}$,
with $j=1,...,k$, the state
\begin{equation}
|\psi_j\rangle = \! \sum_{\sigma,\mathbf{m}} A^{[\sigma]} A^{[m_0]} 
\! ... A^{[m_{n-1}]} A^{[m_n]}_j A^{[m_{n+1}]} 
\! ... A^{[m_{N}]} |\sigma\rangle | \mathbf{m} \rangle ,
\end{equation}
describes one of the $k$ lowest-energy eigenstate of the specified
Wilson chain Hamiltonian; the state corresponding to $j=1$ targets
the ground state. The optimization procedure then works as follows: we
generate a local Krylov space on site $n$ by subsequent application of
the Hamiltonian on each of the $k$ orthonormal states associated with
the array $\mathbf{A}^{[m_n]}$. The resulting Hamiltonian $\hat{H}_n$
has a block structure in the Krylov space, with nonzero elements in
form of $k\times k$ blocks along the diagonal and the first
off-diagonal. Next, we diagonalize the Hamiltonian in the Krylov
subspace and construct from its eigenvectors an updated version of the
array $\mathbf{A}^{[m_n]}$, each element being orthonormal to the
others by construction. To move the orthonormal center of the MPS to
the next site ($n+1$), we form the reduced density matrix
$\rho^{\rm{red},j}_{n,n+1}$ of each component $j$ by tracing out the
rest of the chain and sum them up to form
$\rho_{n,n+1}^{\rm{red}}$. Similar to the original DMRG formulation,
we then diagonalize $\rho_{n,n+1}^{\rm{red}}$, keep only the $D$
largest eigenvalues and use the resulting isometry to move the
orthonormal center to site $n+1$. We repeat the optimization
procedure, sweeping multiple times through the entire chain. 
Convergence was assumed  when the change in each
energy level $E_j$ dropped below the threshold
  $|\delta E_j| < 10^{-11}$, which for our longest multi-level chains
  corresponded to  $\simeq 10^{-3}$ 
of the hopping matrix element $t_N$ to the last site.
In all mVMPS calculations we used bond dimensions of $D=100$,
$d_{n} = 40$.

\begin{figure}[h!]
  \centering
\includegraphics[width=.99\linewidth]{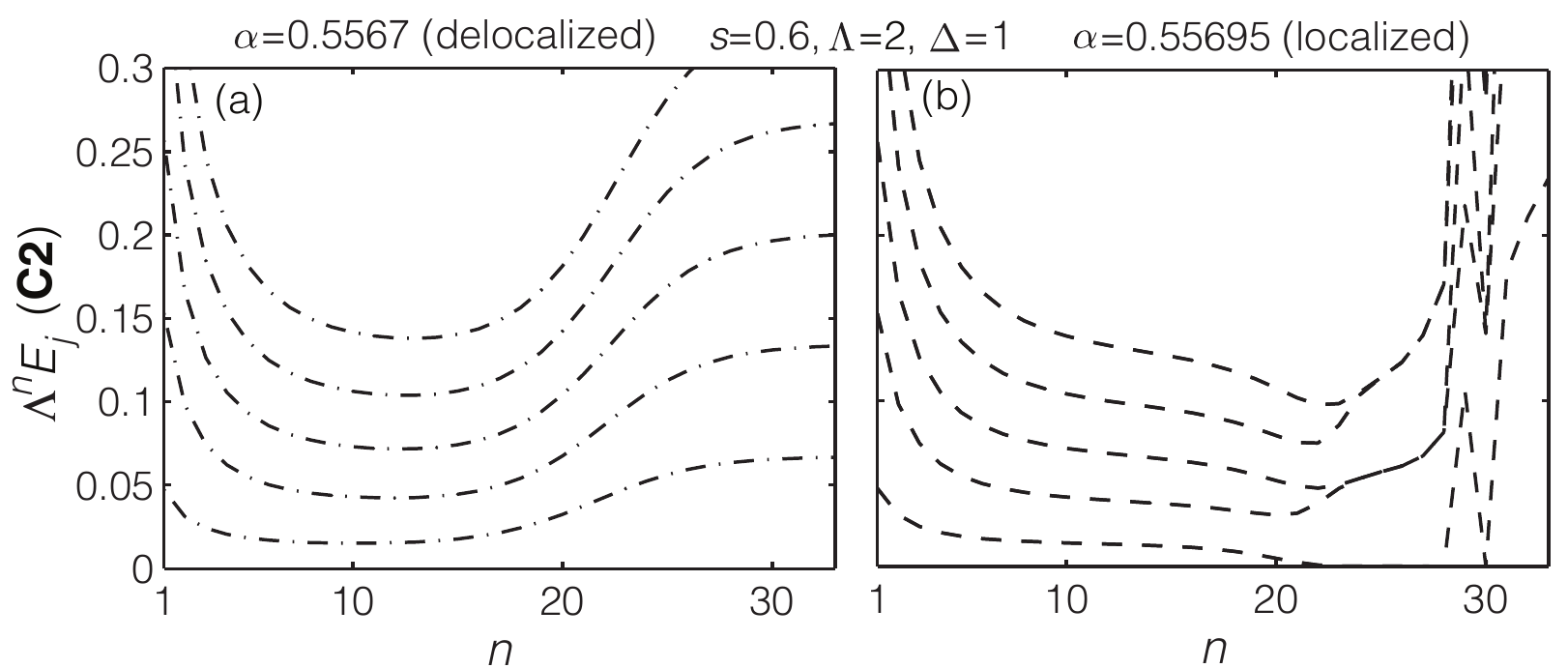} \vspace{-6mm}
\caption{Energy-flow diagrams obtained with mVMPS for the SBM on a
  C2-RWC. (a) Flow to the delocalized fixed point, characterized by a
  unique ground state. (b) Flow to the localized fixed
  point, featuring a doubly degenerated ground-state level. Note that
  the flow gets distorted deep in the localized regime. This is a
  signature of the exponentially growing oscillator shifts which
  cannot be properly dealt with in the mVMPS setup.  }
  \label{Fig.mVMPSflow}
\end{figure}

To account for the mass flow in the energy-level flow of a length-$N$
RWC system, we conduct a \emph{separate} mVMPS calculation for
\emph{every} chain length $N'<N$. This ensures that the $k$
excited states properly take into account the fast- and slow-mode
correction at a particular energy scale, which is crucial for
correctly describing the critical energy flow at a Gaussian
fixed-point. Combining the results for various lengths and rescaling
each set of energies appropriately by a factor $\Lambda^{N'}$, we
obtain the energy-flow diagrams in a variational setup.

In addition to the critical fixed-point flows shown in \Fig{fig:xSBM}
of the main text, we here present the energy-flow to the stable
fixed points in \Fig{Fig.mVMPSflow}. Panel (a) displays the energy
flow to the delocalized fixed point ($\alpha < \alpha_c$), which
features a nondegenerate ground state. In contrast, the fixed point
flow to the localized fixed point ($\alpha > \alpha_c$) in panel (b)
clearly shows a doubly degenerated ground state before getting
numerically distorted by the exponentially growing oscillator
displacements.

The main goal of our mVMPS calculations was to study the
\textit{critical} energy-level flow for the \SBM. Since at the
critical point the truncation of the bosonic Hilbert space is not
problematic, it was not necessary to incorporate the OBB scheme in our
mVMPS setup. Doing so would become essential, however, when studying
the effects of a local bias, $\epsilon \neq 0$, since then
$\langle \hat \sigma_z \rangle \neq 0$. In particular, this would be
needed if one wishes to compute the static susceptibility $\chi(T)$
using not just the VMPS ground-state expectation value for a
length-$N_T$ RWC (as described above), but a thermal average over a
shell of low-lying VMPS eigenstates (as done in NRG). We have
refrained from attempting such combined mVMPS+OBB computations of
$\chi(T)$, since they are numerically expensive, and the
ground-state-based scheme worked very well.

\section{Dissipative harmonic oscillator} \label{sec:DHO}

In this section, we perform a systematic study of the properties of
RWCs in the context of the exactly solvable DHO, which was briefly
introduced in the main text. We compare the RWC and 
SWC setups in detail with respect to the following issues: iteration
details, static susceptibility, and critical coupling $\alphac$.

\begin{center}
\begin{figure*}
	\includegraphics[width=1\linewidth]{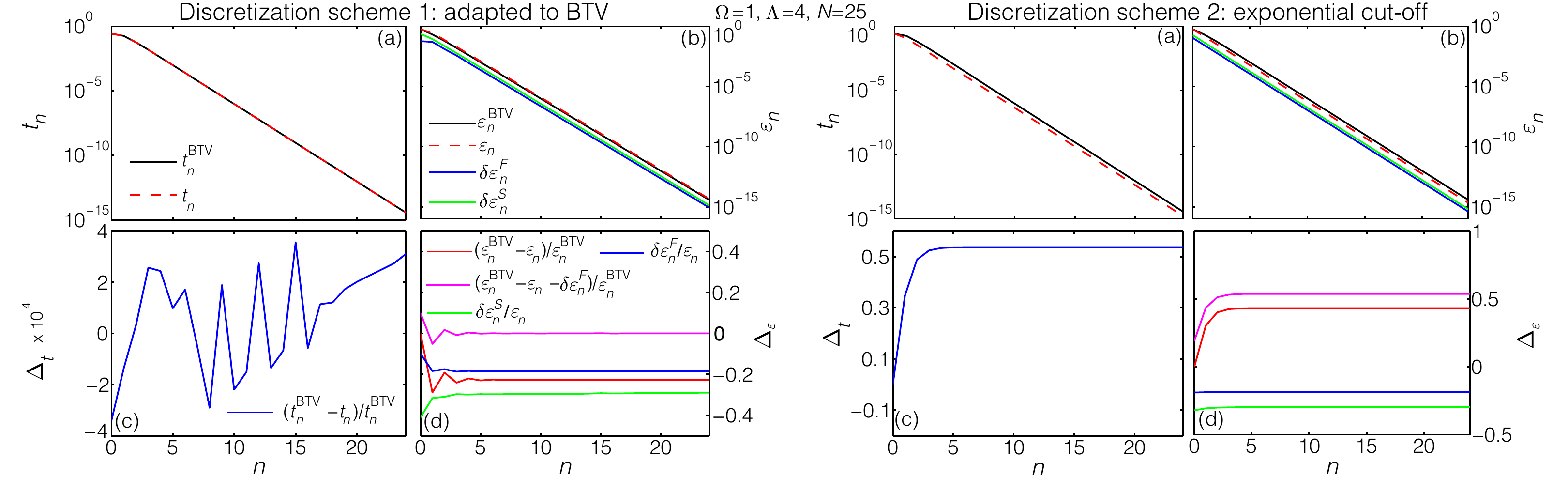}
        \caption{Iteration details: chain parameters.  (a-d) Comparison of the Wilson
          chain parameters $t_n$ and $\varepsilon_n$ for
          $\alpha=0.199$, obtained using the standard discretization
          scheme of BTV \cite{Bulla2005} for $\Lambda = 4$, or using
          two versions of the RWC-approach described above: for
          version 1 (left two columns), $\omega_{\slow,n}$ was
          fine-tuned to ensure that $t_n = t_n^\BTV$; for version 2
          (right two columns), we simply chose
          $\omega_{\slow n}^+ = \omega_{\slow n-1}^+/\Lambda$.  (a)
          $t_n^\BTV$ used by BTV (black) and our $t^\slow_n$ (red
          dashed).  (b) The onsite energies $\varepsilon_n^\BTV$
          (black), our C0 onsite energies $\varepsilon_n$ (red
          dashed), and the shifts $| \delta \varepsilon_n^\fast |$ (blue)
          and $|\delta \varepsilon_n^\slow |$ (green).  Evidently, they
          all scale the same way with $n$.  (c) Relative difference
          $\Delta_t=(t_n^\BTV-t^\slow_n)/t^\BTV_n$ in hopping
          elements. The noisy structure seen for version 1 (left, note the amplification factor of $10^4$)
          reflects the $\omega$-discretization grid used to represent
          the bath correlators $\G^X_n(\omega)$ during the OWC
          construction. (d) Relative differences $ \Delta_\varepsilon$
          of various onsite energies:
          $ \Delta^\CZero_\varepsilon =
          (\varepsilon_n^\BTV-\varepsilon_n) /\varepsilon_n^\BTV$
          (red);
          $ \Delta^\COne_\varepsilon =
          (\varepsilon_n^\BTV-\varepsilon_n - \delta
          \varepsilon_n^\fast)/\varepsilon_n^\BTV$
          (purple);
          $\Delta^\fast_\varepsilon = \delta \varepsilon_n^\fast
          /\varepsilon_n$
          (blue); and
          $\Delta^\slow_\varepsilon = \delta \varepsilon_n^\slow
          /\varepsilon_n$
          (green).  For version 1 (left), the relative difference
          between BTV and C0 energies (no shifts) is quite significant
          throughout ($\Delta^\CZero_\varepsilon \simeq 0.2)$. The
          relative difference between BTV and C1 energies (only fast
          shifts) is significant for early iterations, but becomes
          small ($\Delta_\varepsilon^\COne \lesssim 10^{-3}$) once the
          iteration scheme reaches self-similarity.  For version 2
          (right), both $\Delta_\varepsilon^\CZero$ and
          $\Delta_\varepsilon^\COne$ differ significantly from 0.
          Both the fast and last slow mode shifts are comparable in
          magnitude to the bare OWC energies,
          $\mathcal{O}(\Delta_\varepsilon^{\fast/\slow}) = 1$.
 \label{Fig.couplings-epsilon}
 }
 \end{figure*}
\end{center}

\subsection{Iteration details}

We introduced two types of RWCs in the main text: C1 chains, which
include only the fast shifts ($\delta \varepsilon^{\slow}_N = 0$), and
C2 chains, which contain both slow- and fast-mode shift in
\Eq{eq.EnergyShift}. For completeness, we also discuss a third type of
RWC to be called C0 chains, which by definition include no energy
shifts, i.e.\ $\delta \epsilon^{\slow/\fast}_n = 0$ in \Eq{eq.EnergyShift}.

We have explored two versions of the RWC iteration scheme, that differ
only in the choice of the frequencies $\omega^+_{\slow n}$ that define
the intervals $I_n^\slow = [0,\omega^+_{\slow n}]$. For version
1, we chose $\omega^+_{\slow n}$ in such a manner that the resulting
hopping matrix elements $t_n^\slow$ of the OWC agree with those used
by Bulla, Tong and Vojta (BTV), \cite{Bulla2003,*Bulla2005} to be
called $t_n^\BTV$ [Eq.~(13) of Ref.~\onlinecite{Bulla2005}], with
relative error below $10^{-3}$.  (The error could be further
  reduced, if desired, by using a finer frequency grid for
  representing $\Gamma_n(\omega)$, and more accurately fine-tuning the
  numerical integration routine used to evaluate the integral that
  yields $t_n^\slow$.)  For version 2, we used a plain exponential
discretization, $\omega_{\slow n}^+ = \omega_{\BATH n}^+/\Lambda$.
   
A comparison of the resulting $t^\slow_n$, the bare onsite energies
$\varepsilon_n$ and the shifts $\delta \varepsilon^{\fast/\slow}_n$,
is shown in Fig.~\ref{Fig.couplings-epsilon}.  It has two take-home
messages: First, all these quantities scale the same way with $n$ and
are comparable in magnitude
[Figs.~\ref{Fig.couplings-epsilon}(a,b)]. In particular, the fast and
slow shifts $\delta \varepsilon_n^{\fast/\slow}$ are comparable to the
bare OWC energies $\varepsilon_n$. Second
[Figs.~\ref{Fig.couplings-epsilon}(d)], our RWT energies, both
$\varepsilon_n + \delta \varepsilon_n^\fast$ and
$\varepsilon_n + \delta \varepsilon_n^\fast + \delta
\varepsilon_n^\slow$,
are in general different from the SWC onsite energies
$\varepsilon_n^\BTV$ obtained by BTV using standard Wilsonian
discretization and tridiagonalization, the relative difference being
$\mathcal{O}(1)$. For version 1, however, we note that the relative
difference between $\varepsilon_n + \delta \varepsilon_n^\fast$ and
$\varepsilon_n^\BTV$ becomes negligible for after a few iterations,
but for early ones the difference remains. 

Note that we also explored a third discretization scheme similar to
version 1, with the difference that we fixed the truncation energies
$\omega^+_{\slow n}$ such that the resulting hoppings agree with those
resulting from the improved logarithmic discretization recently
proposed by Zitko and Pruschke (ZP) \cite{Zitko2009}. This leads to
results qualitatively similar to those of version 1, therefore we
refrained from including them in the discussion above.

The results in the main text were obtained using version 2.  This
discretization scheme is particularly appealing due to its accuracy
and simplicity. It is more accurate than standard Wilsonian
discretization, since by construction it reproduces the hybridization
function correctly. The discretization scheme of ZP was devised to
achieve this, too, but our scheme turns out to be more accurate, due
to its inclusion of TBMs (compare green and red symbols in \Fig{Fig.SuscOverview} below). Our discretization scheme is also simpler to implement
than that of ZP, since their chain parameters are found by solving a
differential equation, whereas our chain parameters (fixed fully by
the energies $\tilde \varepsilon_n$ and couplings $t_n^\slow$) 
are
found purely by numerical integrations. The accuracy of the latter can
be easily controlled by distributing the grid points logarithmically
and, in particular, increasing the resolution around the cut-off
frequencies $\omega_{Sn^+}$.  Note that our discretization scheme
offers great flexibility, as one can easily relax the logarithmic
discretization in favour of a linear or mixed one (log-linear or
linear-log) if high- or low-energy properties need to be taken into
account more carefully \cite{Weichselbaum2009}. (The resulting chain
would then have to be treated purely with VMPS methods.)

\begin{center}
\begin{figure}
	\includegraphics[width=1\linewidth]{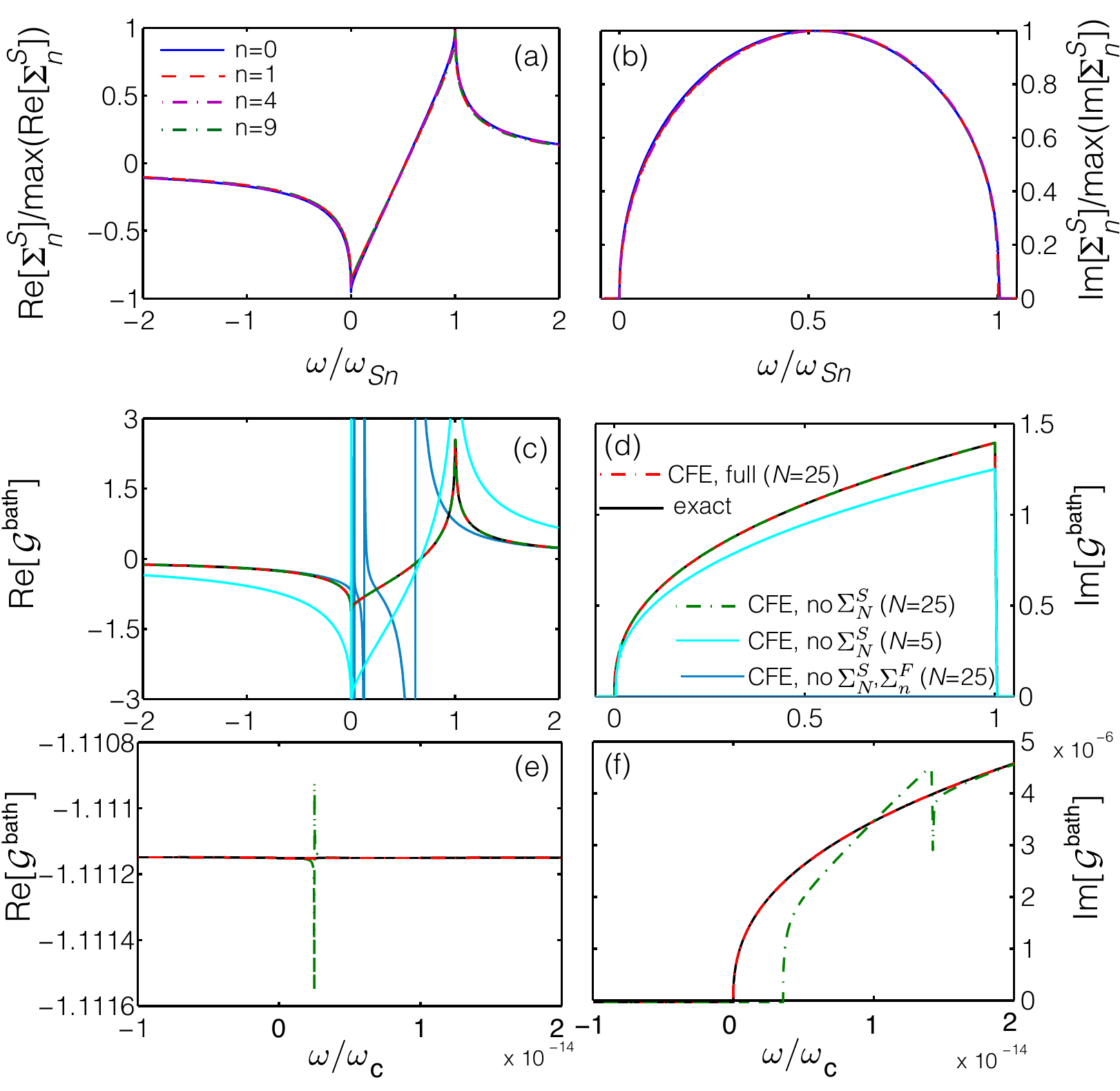}
        \caption{ Iteration details: self-energies. 
(a) ${\rm Re}[\selfE^\slow_n(\omega)]$ and
      (b) ${\rm Im}[\selfE^\slow_n(\omega)]$, plotted vs.\
      $\omega/\omega^+_{\slow n}$ for $n = 0,1,4,9$ (different
      colors), showing that the spectral functions and
      self-energies have a self-similar structure.  (c-f) Various 
CFE representations of $\G^\bath = \G_0$. (c)
      ${\rm Re}[\G^\bath (\omega)]$ and (d) ${\rm Im}[\G^\bath(\omega)]$
      vs. $\omega/\omega_\cut$, calculated directly from
      $\Gamma^\bath(\omega)$ (solid black), or from a CFE
      while including both $\Sigma^\fast_n(\omega)$ and $\Sigma^\slow_N(\omega)$
            with $N\!=\!25$ (dashed red), only $\Sigma^\fast_n(\omega)$ 
            with $N\!=\!5$
      (cyan) or $N\!=\!25$ (dashed green), or neither of the two 
      with $N\!=\!25$ (blue). In
      the latter case, the absence of any imaginary parts in the
      CFE causes ${\rm Im}[\G^\bath(\omega)]$ to vanish and
      ${\rm Re}[\G^\bath (\omega)]$ to have divergences.  Behavior of
      (e) ${\rm Re}[\G^\bath(\omega)]$ and (f)
      ${\rm Im}[\G^\bath(\omega)]$ for $\omega \to 0$ with the same 
      color code as
      in (c,d).  The missing slow-mode term in the CFE using only
      $\Sigma^\fast_n(\omega)$ (green) causes discrepancies for
both the imaginary and the
      real part only in the vicinity of $\omega=0$ illustrating
that the effect of slow-mode shifts becomes noticeable only at
the lowest energy scale of a Wilson chain, associated with its last site.
 \label{Fig.gamma}
}
 \end{figure}
\end{center}

In addition, we have also examined the retarded self-energies
$\selfE_n^\slow$ generated in different iterations $n$ and
checked to what extent our chain parameters reproduce the original
bath correlator $\G^\bath$ [\Fig{Fig.gamma}]. (In this context,
the two discretization schemes yield qualitatively similar results, so
that \Fig{Fig.gamma} only displays version 2.) The main conclusion
drawn from the real and imaginary part of $\selfE_n^\slow$
[Figs.~\ref{Fig.gamma}(a,b)] for the power-law coupling spectrum
$\Gamma^\bathspectrum$ considered here is that the iteration scheme
has a self-similar structure, in that the shape of
${\rm Re}[\selfE_n^{\slow}(\omega)]$ and
  ${\rm Im}[\selfE_n^\slow (\omega)]$ vs.\ $\omega/\omega_{\slow n}$
does not change with $n$. Moreover, the continued fraction expansion
of $\G^\bath$ [Figs.~\ref{Fig.gamma}(c-f)] fully reproduces the
original function (black) if both the fast- and last mode
contributions $\Sigma_n^{\fast/\slow}(\omega)$
are included (dashed red), but if these are neglected 
(dashed green, cyan, blue),
the low-frequency behavior changes significantly.

\subsection{Various averaging schemes} \label{sec:average}

For the VMPS calculations of $\chi(T)$ reported
in the main text, we mimicked thermal averages by ground-state
expectation values of C2-RWCs of length $N_T$. However, we have also
explored several other  averaging schemes. For the sake of
completeness, we briefly describe them here, and in the next section
compare their results for the susceptibility of the DHO.

For a RWC of specified length $N$, we
have explored the following ways of calculating ground-state expectation
values or thermal averages, distinguished by
combinations of the following
labels:
W stands for \textit{Wilsonian} NRG with energy-based truncation;
V for \textit{variational} MPS; \hideandshow{P for a Schmidt projection of a long chain to a
shorter one;} G for a ground-state expectation value; and T for a
thermal average.
For Wilsonian NRG calculations, we denote the eigenstates and
-energies of Wilson shell $n$ by $|s\rangle^\WW_n$ and $E_{sn}^\WW$,
and by $|\GG\rangle^\WW_n$ and $E_{\GG n}^\WW$ for that shell's ground state.
For VMPS calculations, we variationally minimize the ground-state
expectation value of a length-$N$ RWC in the space of all MPS having
specified matrix dimensions. Call the resulting ground state
$|\GG \rangle^\VV_N$, with energy $E_{\GG N}^\VV$.
\hideandshow{For $N'<N$, let $|G\rangle^\VV_N = \sum_{lr}\psi_{lr}
|l\rangle_{N'}^\PP |r \rangle_{N'}^\PP$ be its Schmidt
decomposition in terms of two sets of MPS, $|l\rangle_{N'}^\PP$
and $ |r \rangle_{N'}^\PP$, defined on left and right subchains with
sites $\leqslant N'$ and $> N'$, respectively.
The $|l\rangle^\PP_{N'} $ states form a ``projected basis'' for the
left subchain, describing those linear superpositions of modes from
sites $\leqslant N'$ that contribute most weight to $|G\rangle_N^\VV$.  Let
$|M\rangle^\PP_{N'}$ be the ``maximal weight'' state from this basis
that has largest weight $\sum_{r} |\psi_{lr}|^2$ in $|G\rangle_N^\VV$.
Diagonalizing $\Ham_{N'}^\RWC$ within the projected basis yields a set
of Wilson-shell-like eigenstates and -energies, $|s\rangle_{N'}^\PP$
and $E_{sn}^\PP$. Note that the ground state of this projected shell,
$|G\rangle_{N'}^\PP$, will differ significantly from the maximal
weight state $|M\rangle_{N'}^\PP$ for chains of type C0 and C1, but
not for type C2.}

We write $\langle \hat O \rangle^{\GG Z}_N = {}_N^Z\langle \GG |\hat O
|\GG \rangle_N^Z $ for a ground-state expectation value of type
$Z=\WW$\hideandshow{, $\PP$} or $\VV$.  We write $N_T$ for the length
of a RWC whose smallest excitation energies are comparable to the
temperature,
\begin{eqnarray}
\label{eq:define-T}
\max\{|\tilde \varepsilon_{N_T}|,|t^\slow_{N_T}|\} \simeq T,
\end{eqnarray}
and $\langle \hat O \rangle_{N_T}^{\TT \WW}$ for a thermal average
over all Wilson shell states $|s\rangle_{N_T}^\WW$.%
\hideandshow{ of type $Z=\WW$ or $\PP$.} We will call this
$\TT \WW$-\hideandshow{ or $\TT \PP$-}averaging\hideandshow{,
  respectively}. Thermal averages can also be mimicked using a single
state associated with a length-$N_T$ chain, e.g.\
$\langle \hat O \rangle_{N_T}^{\GG Z} = {}_{N_T}^Z\langle \GG |\hat O
|\GG\rangle_{N_T}^Z $ (GW-\hideandshow{, GP-} or GV-averaging),%
\hideandshow{or
  $\langle \hat O \rangle^{\MM \PP}_{N_T} = {}_{N_T}^\PP\langle M
  |\hat O |M\rangle_{N_T}^\PP $
  (MP-averaging)} because, by the choice of $N_T$, the characteristic
energy spacing for low-energy excitations above such a state is of 
order $T$.  \hideandshow{TP-, GP- and MP-averaging allows a range of
  temperatures to be accessed from a VMPS run for a single length-$N$
  chain to find $|G\rangle_N^\VV$; in contrast, }\mbox{GW-,} TW- and
GV-averaging require calculating a separate length-$N_T$ chain for
every temperature.

\subsection{Susceptibility}
In this section, we compare the various types of RWCs discussed above (C0,C1,C2) and the various averaging schemes by using them to calculate the static susceptibility of the DHO. It is defined by
\begin{equation}
\chi_\stat (T) = \lim_{\epsilon \to 0}
\frac{d\langle a+a^\dagger\rangle_T}{d\epsilon} \ ,
\label{eq:chi-stat}
\end{equation}
where $\langle \dots \rangle_T$ denotes a thermal expectation value.
Its form is easily found analytically \cite{Vojta2010},
\begin{equation}
\chi_\exact (T) =\frac{1}{\Omega+{\rm Re}({\G^\bath(\omega=0)})} \ ,
\label{eq.Susc-exact}
\end{equation}
which, importantly, is  independent of temperature.

Alternatively, the static susceptibility can
also be calculated via the dynamical correlation function
\begin{equation}
C(\omega)=\frac{1}{2\pi}\int_{-\infty}^{\infty}e^{i\omega t}C(t)\mathrm{d}t \ ,
\end{equation}
where $C(t)=\frac{1}{2}\langle[(a+a^\dagger)(t),(a+a^\dagger)]\rangle_T$.
The integral
\begin{equation}
\chi_\dyn (T) =4\int_{0}^{\infty}\frac{C(\omega)}{\omega}\mathrm{d}\omega
\label{eq:chi-dyn}
\end{equation}
can analytically be shown to equal the static susceptibility,
$\chi_\stat(T) = \chi_\dyn(T)$, yielding an important consistency
check for numerical calculations. Our Wilsonian NRG calculations
passed this check for all three types of RWC introduced above (C0, C1,
C2), where $\chi_\stat(T)$ was calculated by evaluating
$\langle \dots \rangle_T$ in \Eq{eq:chi-stat} using a Wilson-shell
thermal average $\langle \dots \rangle_{N_T}^{\TT \WW}$, and
$\chi_\dyn(T)$ was calculated using fdm-NRG
\cite{Weichselbaum2007}. This illustrates the internal consistency of
Wilsonian NRG for a given RWC.  However, none of these calculations
reproduce the exact result (\ref{eq.Susc-exact}). In contrast, the
latter \textit{is} reproduced correctly when calculating $\chi_\stat$
using VMPS on chain type C2, as we now discuss in detail.

\begin{figure}
  \centering
  \vspace{10pt}
	\includegraphics[width=1\linewidth]{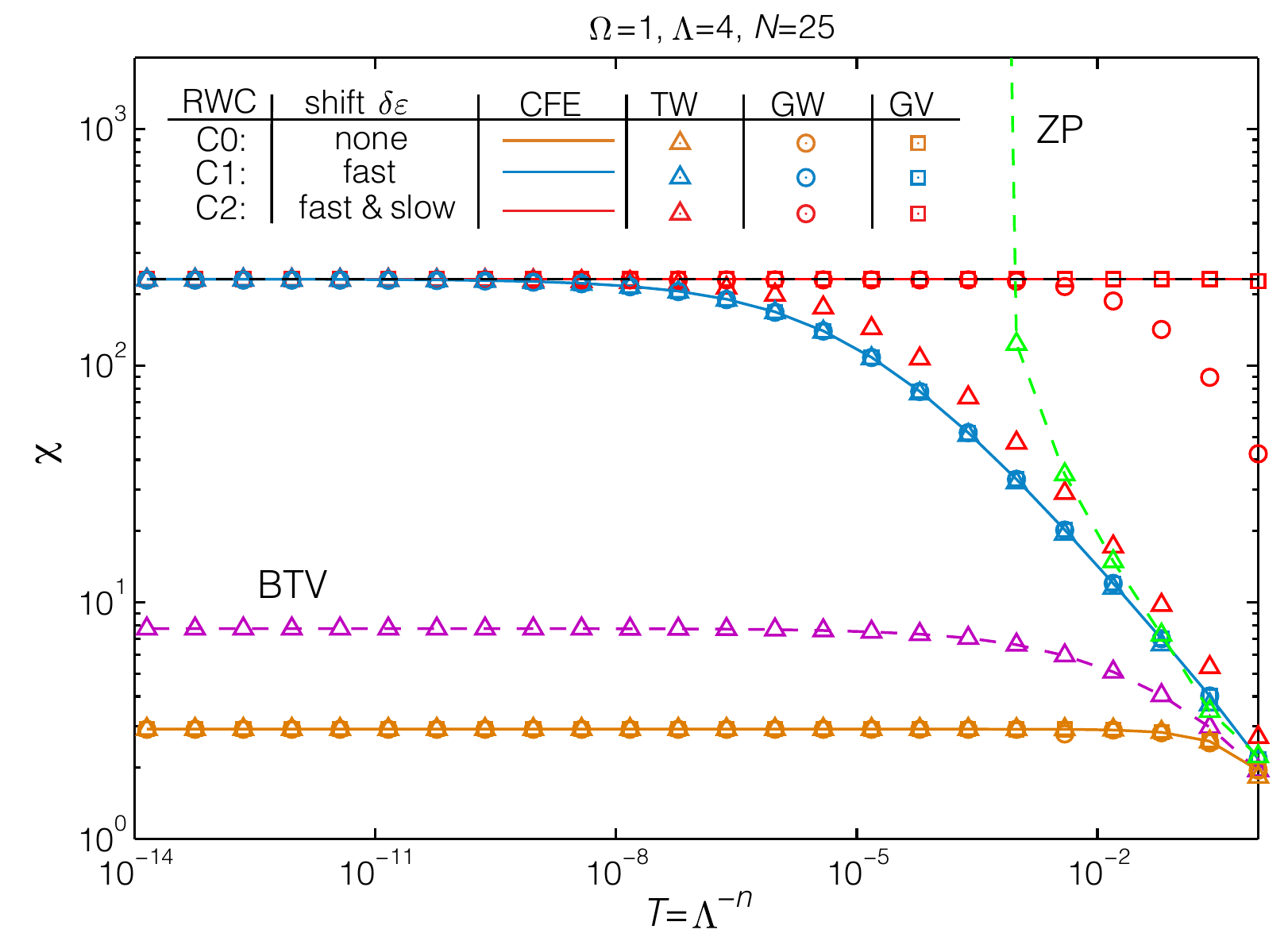}
        \caption{The static susceptibility $\chi_\stat(T)$ of the DHO
          as function of temperature, for $\alpha=0.199$, $s=0.4$. The
          black dashed line gives the exact result $\chi_{\rm exact}$
          expected from Eq.~(\ref{eq.Susc-exact}), the  purple and  green
          symbols the results obtained with standard NRG using
          the discretization scheme of BTV and ZP, respectively.
          The other data are
          numerical results for three types of RWC, C0 (orange), C1
          (blue), and C2 (red), obtained in four different ways.
           The first uses a CFE
          of length $N_T$ to evaluate ${\rm Re}[\G_0 (0)]$ in
          \Eq{eq.Susc-exact} (CFE, solid lines), while including both
          $\Sigma^\slow_n(\omega)$ and $\Sigma^\fast_N(\omega)$ for C2
          (red), only $\Sigma^\fast_n(\omega)$ (blue), or neither of
          the two (orange).  The second evaluates $\langle \dots
          \rangle_T$ in \Eq{eq:chi-stat} as thermal average over
          Wilson shell $N_T$ (TW, triangles). The other two ways
          approximate $\langle \dots \rangle_T$ by an expectation
          value taken w.r.t.\ one of two different single states: the
          ground state $|G\rangle_{N_T}^{\WW}$ of Wilson shell $N_T$
          (\GG \WW, circles) and the variational ground state
          $|G\rangle^\VV_{N_T}$ of a length-$N_T$ chain (\GG \VV,
          squares).  In all
          cases, the derivative $d/d \epsilon$ in \Eq{eq:chi-stat} was
          evaluated numerically,
          using several $\epsilon$-values chosen small enough
          (typically $\ll T$) to ensure that the calculated averages
          depend linearly on $\epsilon$. TW-, GW- and GV-averages
          require a separate run for each combination of $T$ and
          $\epsilon$. }
  \label{Fig.SuscOverview}
\end{figure}

Fig.~\ref{Fig.SuscOverview} shows $\chi_\stat(T)$ for three types of
RWC, C0 (blue), C1 (green), and C2 (red), calculated in
\hideandshow{six}four different ways, involving either a \CFE\ (solid
lines), or a thermal average over Wilson shell $N_T$ (\TT \WW,
triangles), or two\hideandshow{four} types of expectation values
w.r.t.\ states associated with site $N_T$ (\GG \WW, \GG
\VV\hideandshow{, \GG \PP\ and \MM \PP}), as detailed in the figure
caption.  We observe the following salient features.

First, all four\hideandshow{six} methods yield
mutually consistent results both for C0 and for C1, but not for C2
(all orange data lie on a line, as do all blue data, but
not all red data).  Thus the methods differ mainly in
their treatment of slow last modes, which are absent in C0 and C1, but
present in C2.

Second, for C0 (orange), which has the structure of a \textit{standard}
Wilson chain without any TBM information included, $\chi_\Czero(T)$
differs from the exact result, $\chi_\exact$ (dashed black line) in
two important ways: instead of being $T$-independent, $\chi_\Czero
(T)$ increases with decreasing $T$, eventually saturating toward a
constant value, $\chi_\Czero(0)$; and this constant value disagrees
from $\chi_\exact$. The reason for these failings was identified
clearly by VBGA \cite{Vojta2010}: the neglect of TBMs causes ${\rm Re}[\G^\bath
(0)]$ to be represented incorrectly [as is also clearly visible in
Fig.~\ref{Fig.gamma}(c)]. VBGA called the missing contribution
to ${\rm Re}[\G^\bath (0)]$ a ``mass-flow'' error (since near a quantum phase
transition, it implies an artificial scale-dependent shift of the
order-parameter mass).

Third, for C1 (blue), which includes fast but not last slow modes, the
$T$-dependence of $\chi_\Cone(T)$ persists, but its asymptotic
low-temperature value agrees with the exact one,
$\chi_\Cone (0) = \chi_\exact$. Thus, including fast modes is
essential to get the asymptotic value right. Indeed, if they are
  omitted but the slow mode correction included, one obtains curves
  (not shown) whose $T \to 0$ limits corresponds to 
those of C0 curves rather than the exact result.

Fourth, for C2 (red), which includes fast \text{and} last slow modes,
\hideandshow{four} two methods fully reproduce the
$T$-\textit{independent} result $\chi_\Ctwo (T) = \chi_\exact$: CFE
and GV.\hideandshow{, MP and GP. \comment{for MP and GP, this is still
    a guess!}} Their common feature is that both\hideandshow{ all
  four} succeed in fully incorporating the slow-mode contributions to
${\rm Re}[\G^\bath (0)]$. For the CFE this is guaranteed by
construction. For GV-averaging using $|G\rangle_{N_T}^\VV$, it
reflects the ability of the variational MPS scheme to correctly deal
with the large energy shift $\delta \varepsilon_{N_T}^\slow$ at the
end of the length-$N_T$ RWC.

Fifth, the other two methods fail to yield a $T$-independent result
even for C2, since, being based on Wilsonian NRG, they fail to
properly deal with the last slow shift. TW- and GW-averaging involve,
respectively, a thermal average or ground-state expectation value for
Wilson shell $N_T$; but the slow shift $\delta
\varepsilon^\slow_{N_T}$ on the last site is so large that upon adding
it to the chain, some feedback to earlier sites becomes necessary.
Since Wilsonian NRG does not allow for such feedback, while a
variational MPS approach does (through back and forth optimization
sweeps along the chain), TW- and GW-averaging fail, whereas GV-\hideandshow{, GP-
and MP-}averaging does not. We also note that GW does better (yielding a
weaker $T$-dependence) than TW. Presumably the reason is that the
thermal average used by the latter incorporates information from
higher-lying Wilson states $|s\rangle_{N_T}^\WW$, for which the
$\omega = 0$ focus of the static approximation works less well than
for the shell's ground state $|G\rangle_{N_T}^\WW$.

The upshot of the above analysis is that GV-averaging fully meets
  the challenge of correctly computing $\chi(T)$ for the DHO.
  Therefore, GV-averaging was the scheme used for the VMPS calculation
  of $\chi(T)$ reported in 
  Figs.~\ref{Fig.SuscCompare} and \ref{fig:xSBM} of the main text.

\subsection{Critical oupling $\alphac$}
We now turn our attention a small but very important detail
illustrating the power of RWCs to minimize discretization artefacts:
the determination of the critical coupling $\alphac $. Its analytical
value for the parameters used here is $\alphac = 0.2$.  Numerically,
we determined $\alphac$ by monitoring the divergence of the
susceptibility, as described at the end of Sec.~\ref{sec:VMPSm}.

On a SWC with $\Lambda = 4$, the analytical value is either largely
overestimated when using the BTV discretization scheme
($\alpha_\cut^{\rm BTV} \approx 0.228$), or underestimated when using
the improved ZP discretization
($\alpha_\cut^{\rm ZP} \approx 0.1984$); the deviations are due
to the missing information of the TBMs in the Wilson chain
setup. In Fig.~\ref{Fig.SuscOverview}, computed for
  $\alpha= 0.199$, this causes the low-temperature limit of the
  susceptibilities $\chi^{\rm BTV}$ and $\chi^{\rm ZP}$ to lie far
  below or above the analytical value $\chi_\exact$, respectively.
  (In fact, $\chi^{\rm ZP}$ diverges in that figure because
  $\alpha = 0.199$ lies above the critical coupling
  $\alpha_\cut^{\rm ZP}$.)

In contrast, the critical coupling obtained for a C2-RWC matches almost
perfectly with the analytic result. For our setup, we found
$\alpha_c=0.199998$. It is possible to systematically reduce the
deviation from the analytical value of $\alphac$ even further by
improving the resolution of the frequency grid used to represent
$\G_n^X(\omega)$ while constructing a RWC.
Once again, this illustrates the power of our RWC construction and
points out how missing TBMs can introduce systematic
``discretization'' artefacts. Correspondingly, we expect that RWCs
will turn out to be useful for reducing discretization artefacts also
for other dynamic quantities such as local spectral functions.

As $\alpha$ is tuned ever closer to $\alphac$, the VMPS scheme
experiences increasing convergence problems, resulting in increasing
errors for $\chi(0)$.  This is not surprising, because the effective
potential of the DHO becomes ever shallower the nearer $\alpha$
approaches $\alphac$, where the energy of one mode vanishes. That
leads to very large zero-point fluctuations, and a very strong linear
response to small values of $\epsilon$. Increasing the VMPS bond
dimension to keep more states during the calculation failed to
significantly improve $\chi(0)$. We were able to ameliorate this
convergence problem to some extent by implementing an optimized boson
basis designed to incorporate large bosonic displacements. However, as
a matter of principle, this problem will become unmanageable in the
limit $\alpha \to \alphac$.

\FloatBarrier

\section{RG flow towards Gaussian fixed point}
\label{sec:RGflow}

In this section, we connect the numerically obtained energy-level
diagrams to analytical considerations and show that the numerical
results prove the existence of a Gaussian critical fixed point for the SBM
with bath exponents $0<s\leqslant0.5$.

Using a Feynman path-integral representation, the spin-boson model \eqref{hsbm} can be shown to be equivalent -- in the scaling limit -- to the following one-dimensional $\phi^4$ theory:
\begin{equation}
\mathcal{S} = \int \frac{d\w}{2\pi} (m_0 + |\w|^s) |\phi(i\w)|^2 + \int d\tau
\left[u_0 \phi^4(\tau) + \bar{\epsilon} \phi(\tau) \right]
\label{phi4}
\end{equation}
where $\bar{\epsilon}$ is a rescaled bias, and the $|\w|^s$ term
arises from integrating out the oscillator bath with bath exponent
$s$; this generates a bilinear coupling which is long-ranged in
time. $m_0$ is the (bare) mass of the Ising order parameter $\phi$; an
increase of $m_0$ corresponds to a decrease in the dissipation
strength $\alpha$. Finally, $u_0$ is the quartic self-interaction.  By
universality arguments, the same $\phi^4$ theory also describes the
phase transition of a classical Ising chain with $1/r^{s+1}$
interactions.

Power counting in Eq.~\eqref{phi4} yields the scaling dimensions at criticality:
\begin{eqnarray}
\label{scaldim}
{\rm dim}[\phi(\tau)] &=& (1-s)/2 \,,\\
{\rm dim}[u_0] &=& 1-4{\rm dim}[\phi(\tau)] = 2s-1 \,,
\nonumber
\end{eqnarray}
i.e., the system is above (below) its upper-critical dimension for $s<0.5$ ($s>0.5$).

\begin{figure}[t!]
  \centering
\includegraphics[width=.99\linewidth]{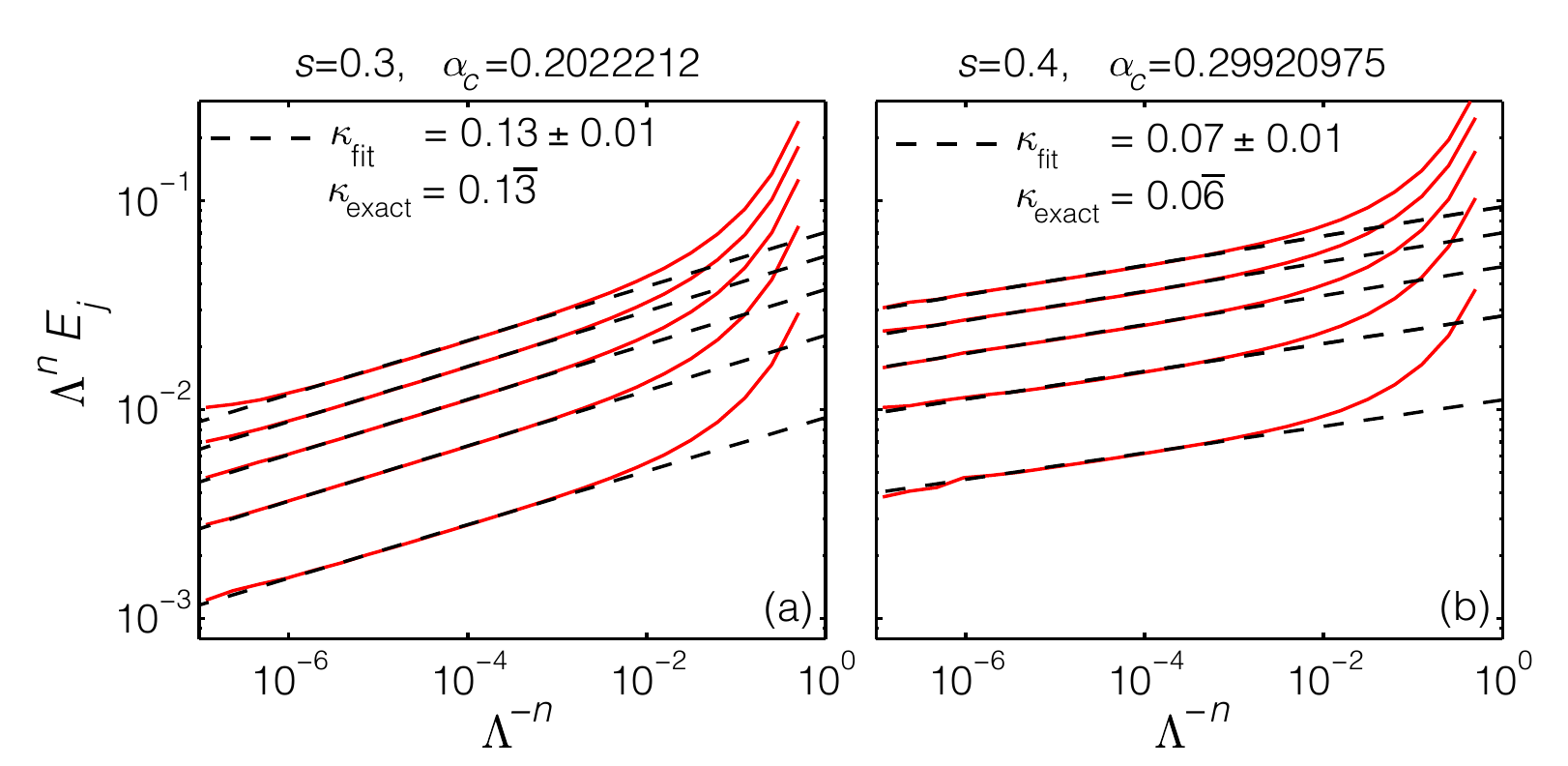}
\caption{Energy-level flow diagram for the SBM on a C2-RWC obtained
  for (a) $s=0.3$ and (b) $s=0.4$ at the critical point. The dashed
  lines illustrate the power-law fits employed to extract the exponent
  $\kappa$ characterizing the Gaussian fixed point. The numerical
  results are in excellent agreement with the analytical prediction
  $\kappa = (1-2s)/3$. 
}
  \label{Fig.KappaFlow}
\end{figure}

In the following, we focus on the regime $0<s\leqslant0.5$ where the
transition is controlled by a Gaussian fixed point.
Although the quartic interaction $u_0$ is irrelevant at criticality,
i.e., its fixed-point value is zero, it is required to stabilize the
system and it influences observables in a nontrivial fashion, hence
it is termed ``dangerously irrelevant''.
The scaling dimension \eqref{scaldim} implies that the leading-order
behavior of the dimensionless renormalized quartic coupling $u$,
defined as $u_0 = \mu^{1-2s} u$ with $\mu$ a renormalization energy
scale, at criticality is given by
\begin{equation}
\label{uscale}
u \propto \varepsilon_\UV^{1-2s}
\end{equation}
with logarithmic corrections present at s=0.5, where $\varepsilon_\UV$
is the running ultraviolet cutoff.  From this we can expect that
the many-body spectrum, i.e., the energy-level flow as described
above, displays families of levels whose spacing goes to zero as the
cutoff energy $\varepsilon_\UV$ goes to zero. This is in contrast to
interacting critical fixed points, here realized for $0.5<s<1$ where
the renormalized $u$ reaches a finite fixed-point value: This causes
the level spacings to approach constant values as
$\varepsilon_\UV\to 0$ (see Ref.~\onlinecite{Lee2005} for a detailed
analysis of NRG fixed-point spectra at interacting critical
points). Both behaviors are nicely borne out by our numerical results
in Figs.~\ref{fig:Gflow} and \ref{Fig.KappaFlow}.

While the effect of $u$ on many observables can be calculated using
(renormalized) perturbation theory, this does not apply to the level
spectrum at criticality: For $u=0$ the spectrum is degenerate
(reflecting a bosonic zero mode), such that the effect of $u$ is
nonperturbative. This zero-mode physics in the presence of a quartic
interaction is captured by the toy-model Hamiltonian for a quartic
oscillator, $\mathcal{H}_4 = p^2/(2m) + u x^4$ in standard
notation. Scaling considerations shows that the eigenenergies of this
model obey the exact scaling $e_i \propto u^{1/3}$. Importantly, this
toy model, if used with a renormalized $u$, describes
\textit{renormalized} energy levels.

Let us now connect the behavior of these renormalized energy levels with
  those generated by NRG. To this end, we note that in NRG the Wilsonian scale
  $\varepsilon_n \propto \Lambda^{-n}$, which is an infrared cutoff,
  controls the RG flow in a way analogous to that of the running UV
  cutoff $\varepsilon_\UV$ in a perturbative RG scheme, as both
  schemes are designed to describe the renormalized physics at the
  scale $\varepsilon_\UV$. Indeed, in an NRG calculation the
  ultraviolet cutoff at a fixed point is a multiple of the infrared cutoff
  $\varepsilon_n$.  Combining the energy scaling of $\mathcal{H}_4$
with Eq.~\eqref{uscale}, we conclude that the low-lying renormalized
energy levels obtained from mVMPS, $\Lambda^n E_j$, scale with the
Wilsonian energy scale $\varepsilon_n \propto \Lambda^{-n}$ as
\begin{equation}
  \Lambda^n E_j \propto \left(\Lambda^{-n}\right)^\kappa~\mbox{with}~\kappa= (1-2s)/3 \, , 
\end{equation} 
characterizing the approach to a Gaussian fixed
point. Fig.~\ref{Fig.KappaFlow} shows a log-log plot of the
energy-level flow for two values of $s$, together with a power-law
fit. We obtain $\kappa=0.13\pm0.01$ for $s=0.3$ and
$\kappa = 0.07\pm0.01$ for $s=0.4$, in excellent agreement with the
analytical prediction, which yields  $0.4/3=0.1\bar3$ and
 $0.2/3=0.0\bar6$, respectively.

\end{document}